\documentstyle[aps,multicol,epsf,pre]{revtex}
\newcommand{\cac} {\varphi_{\rm cac}}
\newcommand{\cmc} {\varphi_{\rm cmc}}
\newcommand{\cD} {{\cal D}}
\newcommand{\cH} {{\cal H}}

\newcommand{\eg} {{\it e.g., }}
\newcommand{\half} {\frac{1}{2}}
\newcommand{\ie} {{\it i.e., }}
\newcommand{\kB} {k_{\rm B}}
\newcommand{\phib} {\varphi_{\rm b}}
\newcommand{\rmd} { {\rm d} }
\newcommand{\rme} { {\rm e} }
\newcommand{\tc} {\tilde{c}}
\newcommand{\tgamma} {\tilde{\gamma}}
\newcommand{\tphi} {\tilde{\varphi}}
\newcommand{\tpsi} {\tilde{\psi}}
\newcommand{\tU} {\tilde{U}}
\newcommand{\tV} {\tilde{V}}
\newcommand{\tW} {\tilde{W}}
\newcommand{\veck} {{\bf k}}
\newcommand{\vecr} {{\bf r}}
\newcommand{\vecx} {{\bf x}}
\newcommand{\vecy} {{\bf y}}

\renewcommand{\theequation}{\thesection.\arabic{equation}}

\begin{document}

%\def\trace{\mathop{\mbox{\large Tr}}}

%-------------------------------------------------------
% 1st page
%-------------------------------------------------------

\draft

\title {Self-Assembly in Mixtures of Polymers and Small Associating
Molecules}

\author {Haim Diamant\footnote{To whom correspondence should be
addressed. Current address:
The James Franck Institute, The University of Chicago,
5640 South Ellis Avenue, Chicago, IL 60637, USA.}
and David Andelman.}

\address{School of Physics and Astronomy,
Raymond and Beverly Sackler Faculty of Exact Sciences\\
Tel Aviv University, Tel Aviv 69978, Israel}

\date{June 13, 2000}

\maketitle

\begin{abstract}
\setlength{\baselineskip}{0.75\baselineskip}

The interaction between a flexible polymer in a good solvent and
smaller associating solute molecules such as amphiphiles
(surfactants) is considered theoretically. Attractive correlations,
induced in the polymer because of the interaction, compete with
intra-chain repulsion and eventually drive a joint self-assembly
of the two species, accompanied by partial collapse of the chain.
Results of the analysis are found to be in good agreement with
experiments on the onset of self-assembly in diverse
polymer--surfactant systems. The threshold concentration for
self-assembly in the mixed system (critical aggregation concentration,
{\it cac}) is always lower than
the one in the polymer-free solution (critical micelle concentration,
{\it cmc}).
Several self-assembly regimes are distinguished, depending on the
effective interaction between the two species. For strong
interaction, corresponding experimentally to oppositely charged
species, the {\it cac} is much lower than the {\it cmc}. It
increases with ionic strength and depends only weakly on polymer
charge. For weak interaction, the {\it cac} is lower but
comparable to the {\it cmc}, and the two are roughly proportional
over a wide range of {\it cmc} values. Association of small
molecules with amphiphilic polymers exhibiting intra-chain
aggregation (polysoaps) is gradual, having no sharp onset.
% --- a
%finding which serves as an important experimental test case for
%the theory.

\setlength{\baselineskip}{1.25\baselineskip}
\end{abstract}

%\pacs{61.25.Hq,82.65.Dp,87.15.Da,61.41.+e}

%---------------------------------------------

%\begin{multicols}{2}

%\pagebreak

\section{Introduction}
%---------------------
\setcounter{equation}{0}

Aqueous solutions containing polymers and smaller solute molecules
are common in biological systems and industrial applications. In
many cases the small molecules are amphiphilic (surfactants) and
may self-assemble with the polymer chains into joint aggregates.
Such systems, synthetic as well as biological, have been the
subject of extensive research in the past few
decades.\cite{review_general1,review_general2}
 The possibility to achieve
polymer--surfactant aggregation using very low surfactant
concentration offers a delicate control over the properties of the
solution, a feature being used in numerous
applications.\cite{review_application}

The current article presents a theory for the onset of
self-assembly in such mixed systems.\cite{letter} The theory
considers the various interactions in a very general way, not
taking into account microscopic details of the small molecules or
the polymer. Indeed, the particular structure of a surfactant may
affect the details of its aggregation. Nevertheless, we suggest
that the {\em onset} of joint polymer--surfactant self-assembly is
mainly determined by simpler, more general considerations.

Self-assembly of polymer--surfactant complexes usually starts at a
well-defined surfactant concentration, the `critical aggregation
concentration' ({\it cac}). One of the most consistent
experimental observations in polymer--surfactant systems is that
the {\it cac} is found to be lower than the `critical micelle
concentration' ({\it cmc}) of the polymer-free surfactant
solution,
\[
  \cac < \cmc.
\]
Consequently, polymer--surfactant systems are commonly divided
into two categories: (i) systems whose {\it cac} is much lower
than the {\it cmc}, $\cac\ll\cmc$; (ii) systems where the {\it
cac} is lower than, but comparable to the {\it cmc},
$\cac\lesssim\cmc$. Experimentally, the former case corresponds to
systems containing a polyelectrolyte and an oppositely charged ionic
surfactant,\cite{review_pe_ionic} \eg polyacrylic acid (PAA) and
dodecyltrimethylammonium bromide (DTAB). The strong electrostatic
attraction between the two species can cause the {\it cac} in such
systems to be several orders of magnitude lower than the {\it
cmc}. The latter case usually corresponds to systems containing a
neutral polymer and an ionic surfactant,\cite{review_neutral_ionic}
 \eg polyethylene oxide (PEO) and sodium dodecylsulfate (SDS).
Somewhat less common are systems containing a polyelectrolyte and
a nonionic
surfactant,\cite{review_Anghel,Vasilescu97,Anghel98} which can be
included in the second category as their {\it cac} is comparable
to the {\it cmc}. Systems where both species are neutral exhibit a
very weak effect.\cite{review_Anghel,review_Lindman,Feitosa96}

The {\it cac} is usually interpreted in terms of the strength of
interaction, or affinity, between the two species. In analogy to
regular micellization,\cite{micelle} $\log(\cac)$ is related to
the free energy of transfer (in units of $\kB T$) of a molecule
from the aqueous solution to a joint aggregate. Evidently, the
affinity should be much stronger for oppositely charged species
(the first category above) than for the other cases, resulting in
a very low {\it cac} in those systems.
%As the bound surfactant is in diffusive contact with
%the surrounding dilute solution,
%$\ln(\cac)$ is inarguably proportional to the free energy of
%binding.
The difficulty, however, is to correctly identify the various
contributions to this free energy. Apart from the bare
interactions among the various molecules, there may be additional
contributions from conformational changes of the polymer
induced by the joint self-assembly. Therefore, construction of a
detailed, reliable molecular model for this complex system is a
complicated task.

Several theories have been presented for polymer--surfactant
aggregation.\cite{review_theory} Most of the models
\cite{micelle_models} attempt to add the polymer as another
ingredient to the already-established thermodynamic theory of
micellization.\cite{micelle} These models are usually applied to
the case of neutral polymers. The prediction of $\cac<\cmc$ does
not arise naturally from the models but depends on the choice of
parameters. Other models,\cite{Shirahama81,Shirahama84} inspired
by the Zimm-Bragg theory of coil--helix transition,\cite{ZimmBragg}
 treat the bound surfactant as an adsorbed
one-dimensional lattice gas. Using two fitting parameters, for the
affinity between the two species and for `binding cooperativity',
they account for binding isotherms in polyelectrolyte--oppositely
charged surfactant solutions. Additional models for
polyelectrolyte--ionic surfactant systems attempt to calculate the
interaction between the two species focusing on electrostatics
\cite{polypeptide,PB_models} and counterion-condensation
effects.\cite{Levin,Colby}

The models mentioned above do not explicitly consider internal
features of the polymer chain. This approach may be justified for
rigid polymers such as DNA or strong polyelectrolytes at low ionic
strength, where electrostatic interactions are not screened. It is
somewhat more questionable in view of the strong conformational
changes observed in flexible polymers upon
self-assembly.\cite{Cabane82,yes_conformation1,yes_conformation2,yes_conformation3}
In fact, most models use various interaction parameters to fit
experimental data, which may implicitly contain conformational
effects (\eg the cooperativity parameter in one-dimensional
models, whose physical origin is unspecified \cite{old_letter}).

Two recent works \cite{recent_models} have treated the polymer
chains in more detail, but in a different context. Both assume
that spherical surfactant micelles have already bound to the
polymer and try to study the additional effect of the adsorbed
chain. In another work \cite{Blankschtein} a detailed molecular
thermodynamic theory of polymer--surfactant
complexes was presented. This model applies to neutral polymers and
contains several molecular parameters.
%Thus, although taking into
%account many effects, it does not provide a clear overall picture
%of the interactions in these mixed systems.

The present work takes a different approach towards the joint
self-assembly of polymers and small solute molecules such as
surfactants. Instead of starting from a model of surfactant
micellization and trying to add the polymer as a further
complication, we rather focus on flexible polymers in solution and
study the effect of small associating molecules, treated as
impurities, on the chain statistics. Unlike surfactant micelles, a
flexible polymer can be treated as a thermodynamic, large system.
Hence, if the polymer undergoes a significant change of
conformation at the onset of self-assembly, then a simple
phenomenological approach might be more successful than in pure
surfactant solutions. We thus conjecture that in a mixed system of
flexible polymers and small molecules the {\it cac} is associated with
a local instability (partial collapse) of the polymer chain. The
instability occurs when attractive correlations induced by the
interaction between the species overcome the intrinsic intra-chain
repulsion. This description is reminiscent of de Gennes' and
Brochard's treatment of a polymer in a binary mixture of good
solvents close to the demixing critical point.\cite{deGennes}
Similar to the latter scenario,
the polymer studied in the current work is
predicted to undergo {\it partial collapse} \cite{deGennes} at the
{\it cac}, which marks the onset of association. The simple
criterion of partial collapse leads to several interesting
predictions which seem to be well supported by experiments.
Furthermore, it allows us to distinguish and explain certain
common, `universal' features in the vast experimental literature
which has accumulated on polymer--surfactant systems.

The theory presented here is phenomenological in nature and does
not consider molecular or structural details. Hence, on one hand,
its results are fairly general, relying on a single requirement
--- that the polymer be flexible enough for its local
conformation to play a significant role in the self-assembly.
(This assumption is quantified in Section~\ref{sec_blob}.) Unlike
detailed molecular models,\cite{Blankschtein} the number of
parameters is reduced to three: one accounting for the affinity
between the two species ($w$), another for intra-chain repulsion
($v$), and the third is the {\it cmc} of the polymer-free
solution, ($\cmc$). On the other hand, the theory is restricted to
the onset of association ({\it cac}) and its vicinity.
Since we are not interested in the micellization itself, we treat
the surfactant solution, as it approaches the {\it cmc}, as a dilute
solution of small associating molecules approaching phase separation.
The theory cannot
provide, therefore, a reliable detailed description of aggregation.
More molecular approaches
can be found in refs~\cite{micelle_models,Blankschtein}.
Nevertheless, it is worth mentioning that models of a simpler,
more general nature were successfully employed in the past for
describing the interaction of polymers with surfactant
monolayers.\cite{polymer_monolayer}

The free energy of the polymer solution is assumed to be
characterized by a single interaction parameter (2nd virial
coefficient). The theory is thus applicable to a dilute as well as
semi-dilute polymer regime. Issues of morphology, phase behavior
and rheology, especially in semi-dilute and concentrated
polymer--surfactant systems, are very interesting and
important,\cite{review_Lindman,review_phase} but lie outside the
scope of the current work.

The structure of the article is as follows. In Section~\ref{sec_cac}
a simple thermodynamic model for the onset of self-assembly in the
mixed system is presented. The main results of this model, as given
in Section \ref{sec_cac_results}, can be
divided into two limiting cases, corresponding to
strong or weak effective interaction between the two species.
In Section~\ref{sec_blob} we present a more refined model,
using a scaling approach to treat the partial collapse of the polymer
in more detail.
We qualitatively discuss in Section \ref{sec_polysoap} the special
case of amphiphilic polymers and polysoaps, which provides
experimental support for our assumptions. Finally, in Section
\ref{sec_summary} we present some conclusions and future
directions.
Throughout the paper we compare our results with available
experiments whenever possible, and stress points where
experimental support is still required. In order to make the
central results as clear as possible, we have put most of the
technical calculations in two appendices. Appendix A contains a
detailed statistical-mechanical calculation, which is used to
verify the general results of Section \ref{sec_cac} while allowing
for their systematic improvement. Appendix B presents the details
of the scaling calculation leading to the results of Section
\ref{sec_blob}.

\section{Thermodynamic Approach}
%-------------------------------
\label{sec_cac}
\setcounter{equation}{0}

\subsection{The Model}
%---------------------

Consider a solution of polymer and smaller solute molecules
whose local concentrations are denoted by $c$ and $\varphi$,
respectively. The free energy density can be divided into three
terms accounting for the polymer contribution, the small solute
one, and the coupling between the two,
\begin{equation}
  f(c,\varphi) = f_{\rm p}(c) + f_{\rm s}(\varphi)
  + f_{\rm ps}(c,\varphi).
\end{equation}
(All energies are expressed hereafter in units of the thermal
energy $k_{\rm B}T$, \ie $f$ has the dimensions of inverse volume.)
We treat the repulsion between monomers of the chains using
a 2nd-virial term,
\begin{equation}
  f_{\rm p} = f_{\rm p}^0 + \half vc^2,
\end{equation}
where $f_{\rm p}^0$ is the free energy of an ideal polymer
solution and $v>0$ is the 2nd virial coefficient (having dimensions
of volume). This treatment is valid for dilute, as well as
semi-dilute polymer solutions.
Since the concentration of both species is
low and we are interested only in the onset of association, the
leading quadratic term in the expansion of $f_{\rm ps}(c,\varphi)$
suffices,
\begin{equation}
  f_{\rm ps} = -wc\varphi,
\label{fps}
\end{equation}
where $w\equiv -\partial^2 f_{\rm ps}/\partial c\partial\varphi$
is a parameter characterizing the interaction strength and having
dimensions of volume.
In fact, as will be shown below, this general model is sufficient
for obtaining our main qualitative results.
However, for the sake of clarity, let us specify an expression
for the small solute contribution as well:
\begin{equation}
  f_{\rm s}(\varphi) = \varphi(\log\varphi-1) - \half u\varphi^2
  - \mu\varphi.
\label{f_s}
\end{equation}
The first term in this expression accounts for the ideal entropy
of mixing of the small molecules, the second describes short-range
attraction, and the third is due to a contact with a reservoir
of small molecules having a chemical potential $\mu$.

%In this section we try to present the most
%general discussion --- apart from the expansion \ref{fps} for
%$f_{\rm ps}$ we do not introduce specific expressions for the free
%energies $f_{\rm s}$ and $f_{\rm p}$. More specific models are
%discussed in Section~\ref{sec_F}.

In the absence of polymer the small solute concentration has a
bulk value, $\varphi=\varphi_{\rm b}$, corresponding to the
minimum of $f_{\rm s}$.
%, where $h_{\rm s}$
%denotes the canonical (Helmholtz) free energy and $\mu$ is a chemical
%potential.
Consider a small perturbation in local concentration,
$\varphi=\varphi_{\rm b}+\delta\varphi$. Assuming that the
solution is both below its {\it cac} and {\it cmc}, $f$ can be
expanded in small $\delta\varphi$ to yield
\begin{equation}
  f = f_{\rm p}(c) + f_{\rm s}(\varphi_{\rm b}) -
  wc(\varphi_{\rm b}+\delta\varphi)
  + \half f_{\rm s}^{''}(\varphi_{\rm b}) \delta\varphi^2,
\label{perturbed_f}
\end{equation}
where $f_{\rm s}^{''}(\varphi)\equiv
\partial^2 f_{\rm s}/\partial\varphi^2$.
In this work we identify the {\it cmc} as the value of $\varphi$
at which, for $c=0$, the solution becomes unstable to
small perturbations, \ie
\begin{equation}
  f_{\rm s}^{''}(\cmc) = 0.
\label{cmc}
\end{equation}

Equation \ref{cmc} is essentially a (spinodal) phase separation
condition. In practice, due to the particular structure of
surfactants and the resulting finite-size effects, the {\it cmc}
does not correspond to a true phase transition,
and $f_{\rm s}^{''}(\cmc)$ is not strictly zero.
In the case of aggregation into finite micelles of typical
aggregation number $n$, a rough estimate for
$\cmc f_{\rm ps}^{''}(\cmc)$ is given by $\varphi_1/\varphi_n$,
the volume-fraction ratio of single molecules and molecules
participating in aggregates. This gives
$\cmc f_{\rm ps}^{''}(\cmc)\sim n^{-1}\rme^{-\epsilon}$, where
$\epsilon$ is the energy per molecule (in units of $k_{\rm B}T$)
gained by aggregation.\cite{micelle}
For typical values of $n\sim 100$ and
$\epsilon$ of a few $k_{\rm B}T$, this is a small, yet finite
number.
Since we are interested in the onset of the joint self-assembly
(which is subsequently found to occur at a lower concentration than
the polymer-free surfactant micellization),
we allow ourselves to ignore these delicate considerations.
Using eq~\ref{cmc}, we thus assume
that for $\varphi\leq\cac$ specific
features of the surfactant can be incorporated in the
phenomenological parameter $\cmc$.
%whose relation to the
%phenomenological free energy, $f_{\rm s}$, is determined by an
%instability condition \ref{cmc}.

In the presence of the polymer, minimization of
eq~\ref{perturbed_f} with respect to $\delta\varphi$ gives
\begin{eqnarray}
%  \delta\varphi &=& \frac {w} {f_{\rm s}^{''}(\varphi_{\rm b})}
%  c
%  \nonumber\\
  f &=& f_{\rm p}^0 + f_{\rm s}(\varphi_{\rm b})
  - wc\varphi_{\rm b} + \half\left( v
  - \frac {w^2} {f_{\rm s}^{''}(\varphi_{\rm b})} \right) c^2.
\label{fmin}
\end{eqnarray}
The last term in $f$ implies
an effective reduction in the 2nd virial coefficient
of the polymer,
\begin{equation}
  v_{\rm eff} = v - v_{\rm ps};\ \ \
  v_{\rm ps} \equiv \frac {w^2} {f_{\rm s}^{''}(\varphi_{\rm b})}.
\label{veff}
\end{equation}
Thus, letting the distribution of small molecules, $\varphi$,
reach equilibrium
%({\it annealing} their degrees of freedom)
has led to an effective attraction between chain
monomers.\cite{annealed}

The polymer will become
unstable when $v_{\rm eff}=0$. At this point $c$ is expected to
increase significantly (due to contraction of chain conformation),
leading to a sharp increase in $\delta\varphi$ as well. We
identify this point, therefore, as the {\it cac}.
Setting $v_{\rm eff}=0$ in eq~\ref{veff} and using eq~\ref{f_s}
for $f_{\rm s}$, we find
the following expression for the {\it cac}:
\begin{eqnarray}
  \cac &=& \cmc F\left( \frac{v}{w^2\cmc} \right) \
  < \ \cmc
\nonumber\\
  F(x) &=& \frac{1}{1+1/x}
  \simeq \left\{
  \begin{array}{ll}
  x - x^2 & \ \ \ x\ll 1 \\
  1 - 1/x & \ \ \ x\gg 1
  \end{array}
  \right.
\label{cac}
\end{eqnarray}
This simple calculation demonstrates the physics governing the
mixed system: the affinity between the flexible polymer and the
small solute induces attractive correlations between monomers,
which compete with the bare monomer-monomer repulsion. The
correlations become stronger as the {\it cmc} is approached, and
{\em they are bound to win before reaching the cmc, \ie
$\cac<\cmc$}. The fact that the {\it cac} is lower than the {\it
cmc} has been established by numerous
experiments.\cite{review_general1,review_general2}
According to the description given here, this
fact is a manifestation of a general effect of equilibrated
(annealed) impurities.

It is important to note that the qualitative features of expression
\ref{cac}, relating the {\it cac} and {\it cmc},
do not depend on the specific model taken for the
small molecules, \ie the expression for $f_{\rm s}$.
In Appendix~A we present a more detailed statistical-mechanical
calculation, yielding eq~\ref{cac} as a first order in an expansion.
Going beyond this first, mean-field approximation gives the
same qualitative relation between $\cac$ and $\cmc$ as in eq~\ref{cac},
with merely a modified function $F$.
This modified function can be written in a close form using the inverse
function $x=F^{-1}(y)$ of $y=F(x)$:
\begin{eqnarray}
  F^{-1}(y) &=& \frac{y(1-y+y^2)}{(1-y)^3}
\nonumber\\
  F(x) &\simeq& \left\{
  \begin{array}{ll}
  x - 2x^2 & \ \ \ x\ll 1 \\
  1 - 1/x^{1/3} & \ \ \ x\gg 1
  \end{array}
  \right.
\label{F_beyond}
\end{eqnarray}
Both expressions for $F(x)$, eqs \ref{cac} and
\ref{F_beyond}, have the same limiting behavior,
\ie $F(x)\simeq x$
for small $x$ and $F(x)\simeq 1$ for large $x$.
The two expressions differ, however, in higher orders. The difference
is particularly significant in the asymptotic approach towards
saturation ($x\gg 1$). The mean-field calculation gives a $x^{-1}$
dependence, whereas the improved analysis yields a much slower trend
towards saturation of $x^{-1/3}$. The difference is also evident in
Figure~\ref{fig_F}, which shows the two results for $F(x)$. Indeed,
large values of the argument $x$ correspond to solute concentrations
approaching the {\it cmc}, $\cac\sim\cmc$,
where solute-solute correlations become
strong and the mean-field approximation should give poor results.

In fact, the leading asymptotic behavior of the function $F$, relating
the {\it cac} and {\it cmc}, can be obtained on very general grounds,
without specifying an expression for $f_{\rm s}$.
To this end we use
the following mathematical construction.
(The uninterested reader can skip the following derivation and
just recall the general result, eq~\ref{cac_general}.)
Let $F$ be a dimensionless
function, such that $\varphi=\cmc F(x)$ solves the equation
%
%\begin{equation}
  $\cmc f_{\rm s}^{''}(\varphi) = 1/x$.
%\label{F_definition}
%\end{equation}
%
($x$ is now merely an unspecified argument.)
According to eq~\ref{cmc}, for $x\rightarrow\infty$ the
solution to the equation is $\varphi=\cmc$. Hence we get the
asymptotic behavior for small arguments, $F(x\gg 1)\simeq 1$.
In the other limit one has $x\rightarrow 0$ and $f_{\rm
s}^{''}(\varphi)\rightarrow\infty$. Since $f_{\rm s}(\varphi)$ is
a well-behaved function for $\varphi>0$, the solution for
$\varphi$ must tend to zero. Hence $F(x\rightarrow 0)\rightarrow 0$.
Moreover, in this limit the solution $\varphi\rightarrow 0$ must
become independent of the fixed parameter $\cmc$, which leads to
the asymptotic behavior $F(x\ll 1)\sim x$.
The general expression for the {\it cac} is thus
\begin{eqnarray}
  \cac &=& \cmc F\left( \frac{v}{w^2\cmc} \right) \
  < \ \cmc \nonumber\\
  F(x) &\sim& \left\{
  \begin{array}{ll}
  x  & \ \ \ x\ll 1 \\
  1  & \ \ \ x\gg 1
  \end{array}
  \right.
\label{cac_general}
\end{eqnarray}

\subsection{Results}
%-------------------
\label{sec_cac_results}

The argument $x=v/(w^2\cmc)$ in eq~\ref{cac_general} determines
the strength of effective interaction between the polymer and the
small molecules. Two limiting cases arise: (i) strong effective
interaction ($x\ll 1$), where $\cac\ll\cmc$; (ii) weak interaction
($x\gg 1$), where $\cac\lesssim\cmc$. The two
limiting behaviors, together with a third one corresponding to
polysoaps (Section~\ref{sec_polysoap}), 
are presented in the diagram of Figure~\ref{fig_diag}.
Note that the distinction
between strong and weak interaction involves not only the bare
interaction between the species, as compared to the interaction
among small molecules, but also intra-chain features.
In our opinion, this observation was not
given proper attention by previous studies.

\subsubsection{Strong Interaction}
%---------------------------------

In the case of strong effective interaction between the two
species, $w^2\gg v/\cmc$ (upper part of the diagram in
Figure~\ref{fig_diag}), the attraction among the small molecules
has no effect on the {\it cac} and, according to eq~\ref{cac_general},
it becomes independent of the {\it cmc},
\begin{equation}
  \cac \sim v/w^2 \ll \cmc.
\label{cac_strong}
\end{equation}
In practice, this case corresponds to systems containing
oppositely charged species, \eg a polyacid and a cationic
surfactant.\cite{review_pe_ionic} Because of the strong
electrostatic interactions, the {\it cac} in such systems is
usually found to be two to three orders of magnitude lower than
the {\it cmc}. In order for the requirement of polymer flexibility
to be fulfilled, the system must contain additional salt which
would screen the electrostatic interactions on length scales
comparable to those of the induced attractive correlations.

Both $v$ and $w$ are expected to be dominated in such systems by
electrostatics, and, therefore, mainly depend on the polymer
ionization degree, $I$, and salt concentration, $c_{\rm salt}$. A
polyelectrolyte solution is a complicated system by itself,
exhibiting diverse behavior as function of $I$ and
$c_{\rm salt}$.\cite{review_pe}
However, two observations can generally be made:
(i) the monomer--monomer parameter, $v$, should have a stronger
dependence on $I$ than the monomer--small solute one, $w$ (the
simplest dependence would be $v\sim I^2$ and $w \sim I$); (ii)
both $v$ and $w$ should have a similar (increasing) dependence on
the Debye screening length, $\lambda_{\rm D}\sim c_{\rm salt}^{-1/2}$, \ie
a similar decreasing dependence on $c_{\rm salt}$. Consequently,
$\cac\sim v/w^2$ should increase with $c_{\rm salt}$ and, somewhat
more surprisingly, be only weakly dependent on $I$. A model which
is focused on the bare interaction between the species would
necessarily yield a {\em strongly decreasing} dependence of $\cac$
on $I$. The weak dependence on $I$ is a characteristic
result of our approach, which takes into account intra-chain
features. It stems from a competition between two effects that
compensate each other: a mutual affinity effect (increasing $I$
strengthens the attraction between the oppositely charged
species), and an intra-chain effect (larger $I$ implies stronger
intra-chain repulsion).\cite{only_ES}

Apart from these rather general conclusions, we may try to reach
more quantitative predictions.\cite{pe_reservation}
 The excluded-volume parameter for a flexible
(weak) polyelectrolyte should roughly scale like $v\sim
I^2\lambda_{\rm D}^2\sim I^2 c_{\rm salt}^{-1}$.\cite{Fixman} (This
result can be simply interpreted as an electrostatic energy
$I^2/\lambda_{\rm D}$ integrated over a volume $\lambda_{\rm D}^3$.) Similarly, we
write for the monomer--small solute parameter $w\sim I c_{\rm
salt}^{-1}$. The resulting {\it cac} should scale, therefore, as
\begin{equation}
  \cac \sim I^s (c_{\rm salt})^t \ \ \ \ \ s=0, \ t=1.
\label{cac_scaling_strong}
\end{equation}
A more detailed calculation, however, yields a different scaling
and is discussed in Section~\ref{sec_blob}
(eq~\ref{cac_scaling_blob}).

Table~\ref{tab_I} summarizes various experimental results for the
dependence of $\cac$ on $I$ in the presence of salt. The first six
experimental systems presented in the table exhibit vanishing
dependence on $I$, the next four --- an increasing dependence, and
the last one --- a weak decrease. The fact that most
experiments found a vanishing or slightly {\em increasing}
dependence of $\cac$ on $I$ clearly indicates the
important role of intra-chain features in the
self-assembly.
If intra-chain features are disregarded, one would expect,
upon increasing $I$, a stronger attraction between the
oppositely charged species and, hence, a sharp {\it decrease}
in the {\it cac}, \ie an opposite trend to the one observed in
Table~\ref{tab_I}.

As mentioned above, the solution must contain enough salt for our
description to hold. Electrostatic interactions should be screened
on length scales comparable to the correlation length in the
surfactant solution, \ie the Debye screening length, $\lambda_{\rm D}$,
should be smaller than a few nanometers. For monovalent salt, it
means that the salt concentration should exceed about 10 mM. This
might explain the inconsistent trends observed in systems
containing only about this amount of salt (Table~\ref{tab_I}).

Figure~\ref{fig_salt} shows experimental results for the
dependence of $\cac$ on $c_{\rm salt}$, taken from eleven different
experiments with different mixtures of polymers, surfactants and
mono-valent salts. All systems exhibit an
increasing dependence on $c_{\rm salt}$, in qualitative agreement
with our finding. Previous works focused on small differences in
the slopes of $\log(\cac)$ {\it vs}.\ $\log c_{\rm salt}$ for different
systems, attributing them to different ionization degrees of the
charged aggregates.\cite{Colby,Malovikova84} While such effects
are probably present, we would rather like
to draw the attention to the striking {\em uniformity} of the
slopes --- all of the graphs in Figure~\ref{fig_salt}, representing
eleven different polymer--surfactant systems, have fitted slopes
in the narrow range of 0.68--0.77, namely $\cac\sim (c_{\rm
salt})^t$ with $t\simeq$0.68--0.77.\cite{multivalent}
This uniformity was not pointed out before.
It might indicate that specific molecular details are not essential
to determining the onset of self-assembly, as suggested here.
Quantitatively, the observed power law disagrees with the exponent
$t=1$ in eq~\ref{cac_scaling_strong}. We return to this point
in Section~\ref{sec_blob}.

It is important to note again that our results hold for flexible
polymers only. A different behavior as function of $c_{\rm salt}$
is observed for stiff polymers such as DNA or
proteins.\cite{review_Ananth}
Similarly, the {\it cac} in salt-free
solutions of strongly charged polyelectrolytes, which cannot be
regarded as flexible chains, depends sensitively on
$I$.\cite{rigid_pe}

\subsubsection{Weak Interaction}
%-------------------------------

In the other limiting case of eq~\ref{cac_general},
$w^2\ll v/\cmc$ (lower part of the diagram in
Figure~\ref{fig_diag}),
the effective interaction between the polymer and small molecules
is weak and the {\it cac} and {\it cmc} become comparable
(yet still $\cac<\cmc$),
\begin{equation}
  \cac = A\cmc,
\label{cac_weak}
\end{equation}
where $A=F[v/(w^2\cmc)]\lesssim 1$ can be considered essentially
as a prefactor which is not
very sensitive to changes in $v$, $w$ or $\cmc$
[since $F(x)$ is close to saturation; cf. eq~\ref{cac_general}].
Experimentally, this weak-interaction limit
applies to systems where at least one of the species is uncharged,
\eg neutral polymers interacting with ionic surfactants
\cite{review_neutral_ionic} or polyelectrolytes interacting with
nonionic surfactants.\cite{Vasilescu97,Anghel98} The {\it cac} is
expected to depend in this case on molecular details. However,
{\it most of this complicated dependence is incorporated in
$\cmc$ itself}. In other words, changing various parameters (\eg
ionic strength) may lead to considerable changes in both the {\it
cmc} and {\it cac}; yet, according to eq~\ref{cac_weak},
their ratio is expected to remain roughly
constant.
We note again that the model is not presumed to properly account
for the {\it cmc} itself. It is expected, however, to correctly capture
the relation between the {\it cac} and {\it cmc}, due to the
particular behavior of the polymer at the {\it cac}.

The simple prediction given in eq~\ref{cac_weak} is verified
in various experiments, as summarized in Table~\ref{tab_weak}.
In each of the four experimental systems presented in
Table~\ref{tab_weak}, the ratio $\cac/\cmc$ remains roughly
constant, sometimes over a wide range of {\it
cmc} values.\cite{depend_cmc}

\section{Scaling Approach}
%-------------------------
\label{sec_blob}
\setcounter{equation}{0}

\subsection{The Model}
%---------------------

The treatment given in Section~\ref{sec_cac} for the onset of
self-assembly is not accurate enough and should be regarded as a
first step in a more rigorous calculation. Its description of the
{\it cac} resembles a `shifted' $\theta$ collapse --- a sharp
transition of polymer conformation occurring when the 2nd virial
coefficient changes sign.
In practice, however, flexible polymers
do not exhibit a sharp coil-to-globule collapse at the {\it cac}.
Their association with small surfactant molecules exhibits a
steep, albeit continuous increase at the {\it cac}, the finite
slope being associated with the
`binding cooperativity'.\cite{Sergeyev}

The difference between a $\theta$ point and the {\it cac} lies in
the different ranges of competing interactions. In a regular
$\theta$ point the competing interactions (between monomers and
between monomers and solvent molecules) have a similar short
range. This leads to a sharp conformational collapse which is
stabilized by three-body interactions (3rd virial coefficient
term). By contrast, in the system discussed here the strong,
short-range repulsion between monomers is overcome by weaker, yet
longer-range attractive correlations.
These attractive correlations are induced by the small associating
molecules interacting with the polymer, as has been found in Section
\ref{sec_cac}.
As a result of the competition between interactions of different
ranges, the polymer
undergoes a more moderate {\it partial collapse}
into sub-units (`blobs'), such that the interaction between
monomers within each blob is dominated by the short-range
repulsion, whereas the interaction between blobs is dominated by
the attractive correlations.

This behavior resembles the one previously discussed by de Gennes
and Brochard for a polymer in a binary mixture close to the critical
demixing point.\cite{deGennes}
An important difference, however, is that the correlation length
in the system of ref~\cite{deGennes} may become arbitrarily large.
The solution discussed here, by contrast, is not close to a critical
point but approaches a point of phase separation or micellization.
Thus, the correlations may become strong but their range remains
finite.

Partial collapse is essentially a `smoothed' $\theta$ transition ---
the rescaled `chain of blobs' is at a $\theta$ point, while
on length scales smaller than the blob size the chain is almost
unperturbed.
Throughout the regime of partial collapse, as small solute molecules
are added, the sub-division of the chain into blobs is adapted so
as to keep the rescaled chain at a $\theta$ point.
Association, thus, progresses {\it
continuously}, as the blobs become smaller and more numerous,
and the local monomer concentration gradually increases.
In the following analysis, the added solute molecules
(\eg surfactants) do not appear explicitly. They are accounted
for via the effective interaction which they
induce in the polymer. This interaction has a typical amplitude,
$e^2$, and a typical range, $\xi$, both of which implicitly depend
on the solute concentration $\varphi$.
Since, for very long chains, $\xi$ is the only length scale in the
problem, it must also be the typical size of a
blob.\cite{deGennes}

Let us consider, therefore, a chain of blobs of size $\xi$, each
containing $g$ statistical (Kuhn) segments, as sketched in
Figure~\ref{fig_blob}.\cite{single_chain}
If each blob contains a large number of segments, its size $\xi$
is related to the number $g$ by a power law,
\begin{equation}
  \xi \sim g^\nu a^{z} v^{(1-z)/3},
\end{equation}
where $a$ is the length of a Kuhn segment. In the case of
excluded-volume repulsion in three dimensions, the Flory argument
yields $\nu=3/5$ and $z=2/5$.\cite{dG_book}
Further properties of the `chain of blobs' can be studied using
scaling arguments, as presented in detail in Appendix~B.
This calculation leads to the following
relations between $g$, $\xi$ and the phenomenological
parameters introduced in Section~\ref{sec_cac}:
\begin{eqnarray}
  g &\sim& \left(\frac{v}{v_{\rm ps}}\right)^{1/\alpha}
    \left(\frac{a^3}{v}\right)^{z/\alpha}
% \nonumber\\
 \nonumber\\
  \xi &\sim& \left(\frac{v}{v_{\rm ps}}\right)^{\nu/\alpha}
    \left(\frac{a^3}{v}\right)^{\nu z/\alpha}
    a^z v^{(1-z)/3}
% \nonumber\\
 \nonumber\\
  \alpha &\equiv& 2-3\nu,
\label{g_xi}
\end{eqnarray}
where $v_{\rm ps}(\varphi)$ is the effective reduction in the
2nd virial coefficient due to the added solute (surfactant),
defined in eq~\ref{veff}.

\subsection{Results}
%-------------------

Several interesting observations arise from eq~\ref{g_xi}. In
order for the results to be consistent, $g$ and $\xi$
must increase with decreasing
$v_{\rm ps}(\varphi)$ so that the entire chain should reduce to
a single blob for small enough $\varphi$. Hence, the
self-consistency condition is
\begin{equation}
  \alpha > 0 \ \ \Longleftrightarrow  \ \ \nu < 2/3.
\label{sc}
\end{equation}
This self-consistency condition gives a precise definition for
the requirement of
polymer flexibility --- on the scale of the correlation length
in the
solution the chain statistics should satisfy $\nu<2/3$. (In
particular, the chain should not be stretched, having $\nu=1$.)
For example, in polyelectrolyte solutions this condition sets a
lower bound for salt concentration, below which the chain would be
too stretched on the length scale of $\xi$, and the
partial-collapse picture described here would become invalid.

Repeating the calculation for
chains embedded in $d$ dimensions, the same result as
eq~\ref{g_xi} is obtained, with $\alpha=2-d\nu$. This
self-consistency condition, $\alpha>0$, is similar to well known
results for the critical behavior of disordered systems.
For both equilibrated (annealed) and frozen (quenched) disorder
--- Fisher renormalization
\cite{Fisher} and the Harris criterion,\cite{Harris} respectively
--- the critical behavior is affected by impurities if $\nu<2/d$,
\ie $\alpha>0$.
% (the so-called {\it crossover exponent}).
Thus, in a similar way,
small solute molecules affect the conformational transition of a
polymer if $\nu<2/d$.\cite{d_4}

We stress again that the solution discussed here is not close
to a critical point and, hence, the correlations induced in the
polymer may be strong but their range remains finite.
As a result, the blobs cannot be arbitrarily large, \ie
$g$ and $\xi$ are bounded by certain maximum values, $g^*$ and
$\xi^*$.
Since $\xi^*$ characterizes the range of correlations in
the solution of small molecules (surfactants), it can be estimated
by the typical size of aggregates (micelles) formed at the
{\it cmc}, \ie typically a few nm.
The value of $g^*$, in turn, is given by the number of monomers
in a blob whose size is equal to $\xi^*$.

The onset of association in the mixed system (the {\it cac})
is expected when blobs can form, \ie when
the value of $g$ required for partial collapse (eq~\ref{g_xi})
becomes smaller than the threshold $g^*$.
Setting the right-hand side of eq~\ref{g_xi} for $g$ equal
to $g^*$, and substituting eq~\ref{veff} for $v_{\rm ps}$
and the function $F(x)$ defined in Section~\ref{sec_cac}, we find
the following expression for the {\it cac}:
\begin{eqnarray}
  \cac &=& \cmc F\left[ (g^*)^{-\alpha}
  \left(\frac{a^3}{v}\right)^z
  \frac{v}{w^2\cmc} \right]
\nonumber\\
  F(x) &\sim& \left\{
  \begin{array}{ll}
  x  & \ \ \ x\ll 1 \\
  1  & \ \ \ x\gg 1
  \end{array}
  \right.
\label{cac_blob}
\end{eqnarray}
Comparison to eq~\ref{cac_general} shows that the less refined analysis
of Section~\ref{sec_cac} corresponds, in fact,
to complete collapse ($g=1$), rather than the actual partial
collapse ($g=g^*$).
%and to the simple athermal case of $v=a^3$.

The similarity to the Harris criterion persists.
Suppose that we could somehow control the correlations in the
solution, \ie tune $g^*$, while keeping the concentration of
small molecules
$\varphi$ fixed (this might be achieved, for example,
by changing the temperature).
In such a scenario, instead of $\cac$, there would be a certain
value of $g^*$ corresponding to the onset of self-assembly.
For $\varphi/\cmc\ll
1$ we find from eq~\ref{cac_blob} that this value of $g^*$ satisfies
$g^*\sim\varphi^{-1/\alpha}$.
It implies that in the absence of `impurities' ($\varphi=0$)
only complete collapse of an infinite chain can take place
($g^*\rightarrow\infty$), whereas for finite $\varphi$ a smoother,
partial collapse into finite blobs is possible.
This is analogous
to Harris' result for the broadening of a critical point by
impurities,\cite{Harris} where, instead of a sharp transition
at a critical temperature $T=T_{\rm c}$, there is a smooth
crossover along a range of temperatures $\Delta T$.
Harris' result for this broadening is
$\Delta T/T_{\rm c}\sim\varphi^{1/\alpha}$,
where $\varphi\ll 1$ is, in this case, the concentration of
impurities.
Recall that
the number of monomers serves as a conjugate variable to
$\Delta T/T_{\rm c}$ in the analogy between polymers and critical
phenomena,\cite{dG_book} \ie $g^*\rightarrow\infty$ corresponds to
$\Delta T/T_{\rm c}\rightarrow 0$.
The smoothing of the $\theta$
collapse of an infinite chain into partial collapse of finite
blobs, due to small solute molecules, is thus analogous to the
smoothing of critical points by impurities.\cite{no_Fisher}

Another result of the partial-collapse picture is that at the {\it
cac}, since the `chain of blobs' is at a $\theta$ point, it should
obey Gaussian statistics. Hence, the radius of gyration of the polymer
should scale with the polymerization degree, $N$, as $N^{1/2}$.
This prediction is still to be confirmed experimentally.
Contraction of the polymer at the {\it cac} was observed in
several systems.\cite{Nilsson,Cabane85,Dawson} Additional support is
found in light-scattering and potentiometric experiments reporting
a surprisingly weak interaction between charged aggregates of
ionic surfactant and neutral polymer.\cite{Reed,Gilanyi81}

In the strong-interaction regime [small argument of $F(x)$ in
eq~\ref{cac_blob}], the partial-collapse analysis leads to an
expression for the {\it cac} which is different from the one given
in Section~\ref{sec_cac} (compare to eq~\ref{cac_strong}),
\begin{equation}
  \cac \sim (g^*)^{-\alpha} \left(\frac{a^3}{v}\right)^z
  \frac{v}{w^2}.
\label{cac_strong_blob}
\end{equation}
In polyelectrolyte systems relevant to this regime, the Kuhn
length $a$ should be taken as the electrostatic persistence
length.\cite{review_pe,pe_reservation}
For flexible, weak polyelectrolytes it depends on the
polymer ionization degree, $I$, and salt concentration, $c_{\rm
salt}$, as $a\sim I\lambda_{\rm D}\sim Ic_{\rm salt}^{-1/2}$.\cite{persistence}
As in Section~\ref{sec_cac}, we take the simple,
weak-polyelectrolyte expressions for $v$ and $w$:\cite{Fixman}
 $v\sim I^2 c_{\rm salt}^{-1}$ and $w\sim I c_{\rm salt}^{-1}$. The
last factor to account for in eq~\ref{cac_strong_blob} is the
threshold number of monomers, $g^*$, whose dependence on $I$ and
$c_{\rm salt}$ is unknown. We consider two simplified cases: (i)
constant threshold for the number of monomers in a blob, $g^*$; (ii)
constant threshold for the spatial size of a blob, $\xi^*\sim (g^*)^\nu
a^z v^{(1-z)/3}$. In reality, neither of these cases is expected
to be strictly correct.
%(Since $\xi^*$ is a property of the surfactant
%solution alone (it is the correlation length in the solution)
%whereas $g^*$ involves also properties of the chain,
%the latter case is probably more accurate.)
The resulting dependence
of the {\it cac} on $I$ and $c_{\rm salt}$ for the two simplified
cases is
\begin{eqnarray}
  \cac &\sim& I^s (c_{\rm salt})^t
\nonumber\\
  s &=& \left\{ \begin{array}{ll}
  z=2/5, & \mbox{constant} \ g^* \\
  z+(\alpha/3\nu)(2+z)=2/3, \ \ & \mbox{constant} \ \xi^*
  \end{array}\right.
\nonumber\\
  t &=& \left\{ \begin{array}{ll}
  1-z/2=4/5, & \mbox{constant} \ g^* \\
  1-z/2-(\alpha/6\nu)(2+z)=2/3, \ \  & \mbox{constant} \ \xi^*
  \end{array}\right.
\label{cac_scaling_blob}
\end{eqnarray}
where we have used again the Flory values $\nu=3/5$ and
$z=2/5$.\cite{dG_book}

Comparison %of eq~\ref{cac_scaling_blob}
to eq~\ref{cac_scaling_strong} shows that the partial-collapse
analysis has led to quantitatively different results. Instead of a
vanishing dependence on $I$ we find a weakly increasing one. Both
vanishing and weakly increasing dependencies were observed
experimentally (Table~\ref{tab_I}). As discussed in
Section~\ref{sec_cac}, these findings qualitatively support our
approach, emphasizing the significance of intra-chain
interactions. In order to quantitatively determine the correct
dependence on $I$, more experiments are needed, in particular at
higher ionic strength.

Equations \ref{cac_scaling_strong} and \ref{cac_scaling_blob}
differ also in the quantitative dependence on $c_{\rm salt}$. The
dependence in eq~\ref{cac_scaling_blob} agrees
with the experimentally observed power laws (Figure~\ref{fig_salt}),
having exponents of $t\simeq$ 0.68--0.77.

\section{Comments on Amphiphilic Polymers \& Polysoaps}
%------------------------------------------------------
\label{sec_polysoap}
\setcounter{equation}{0}

Our basic conjecture, regarding instability of polymer
conformation at the onset of self-assembly, can be qualitatively
supported by considering a special class of polymers ---
associating polymers that form intra-chain aggregates in the
absence of any additional associating solute. A good example for
this case are amphiphilic side-chain polymers, which consist of a
hydrophilic backbone (usually a polyacid) and many hydrophobic
side chains.\cite{Benrraou92,Anthony96,Zana_polysoap} Within a
certain range of hydrophobicity, those polymers exhibit
intra-chain aggregation while still remaining water-soluble, in
which case they are called {\it polysoaps}. By synthesizing
polymers with various side-chain lengths and controlling their
ionization degree, a crossover between regular polyelectrolyte
behavior and intra-chain association (polysoap) can be
observed.\cite{Zana_polysoap}

According to our description, a polysoap is already partially
collapsed. No further instability is supposed to occur upon
addition of small solute molecules and, hence, no sharp onset of
self-assembly is expected. Association of small molecules to such
a chain should progress gradually as function of concentration, by
means of partitioning of molecules between the aqueous solution
and the already-collapsed polymeric aggregates.

The association of ionic surfactants with such hydrophobically
modified polyelectrolytes, poly(maleic acid-co-alkylvinyl ether),
was thoroughly studied.\cite{Benrraou92,Anthony96,Zana_polysoap}
 When the polymer is in the regular polyelectrolyte regime
(\eg having short side chains of 1--4 hydrocarbon groups),
a sharp, cooperative binding is observed. On the other hand, when
the polymer behaves as a polysoap (having longer side chains
and exhibiting intra-chain aggregation),
surfactant association is found to be gradual with
no apparent {\it cac}.\cite{Benrraou92}  We regard this
experimental observation as a strong support for our conjecture,
associating the {\it cac} with partial collapse.

Furthermore, let us consider amphiphilic polymers which still
behave like polyelectrolytes but lie very close to the polysoap
regime. This can be achieved, for example, by tuning their
ionization degree. The effective 2nd virial coefficient of such
polymers should be small, leading, according to
eq~\ref{cac_strong_blob} (or \ref{cac_strong}), to low {\it
cac}. The physical reason is that close to the polysoap regime the
stability of the polymer is only marginal,
\ie $v$ is close to zero even in the absence of additional solute
(surfactant).
Hence, a small amount of
solute is sufficient to cause self-assembly. In this region
of $v\gtrsim 0$, therefore, intra-chain features, rather than the
affinity
between the two species, determine the onset of self-assembly. As
a result, the {\it cac} can be significantly reduced without a
significant change in the bare interaction, or, moreover, even if
the bare affinity becomes {\it weaker} (\eg by reducing $I$).
There are two available experimental works demonstrating this
surprising effect,\cite{Benrraou92,Chen98} as shown in
Figure~\ref{fig_amphi}. Both experiments involved amphiphilic
polyelectrolytes whose charge density was varied. Although
reducing charge density must {\it weaken} the interaction with the
oppositely charged surfactant, the {\it cac} was shown to {\it
decrease}, the effect becoming sharp close to the polysoap limit.

The polymers discussed above have many hydrophobic groups along
their backbone. Also worth mentioning are experiments involving
polyelectrolytes with a very small number of hydrophobic
groups.\cite{Magny94}  In this case too, the {\it cac} was found to
significantly decrease upon increasing the degree of hydrophobic
modification, implying a sensitive dependence on intra-chain
features.

\section{Conclusions}
%--------------------
\label{sec_summary}

Focusing on the onset of self-assembly (the {\it cac}), we have
presented a unified description of the interaction between a
flexible polymer and small associating molecules in dilute
solution. Utilizing a conjecture of partial collapse of the
polymer at the onset of self-assembly, we have obtained simple
predictions which seem to be well supported by experiments on
diverse polymer--surfactant systems.

Apart from the bare interaction between the two species, we argue
that intra-chain interactions have an important role as well. In
certain cases, such as systems involving amphiphilic polymers,
intra-chain features may even become the dominant factor
determining the {\it cac}. The interplay between various
interactions in the system (monomer--solute, monomer--monomer and
solute--solute) leads to three self-assembly scenarios, which are
summarized in the diagram of Figure~\ref{fig_diag}. By modifying
intra-chain features of the polymer, one can obtain a crossover
between the various self-assembly regimes without necessarily
changing the bare interaction between the two associating species.
An interesting experiment would be to take a weakly interacting
system (\eg a polyacid like PAA and a nonionic surfactant like
C$_n$E$_m$) and by modifying the polymer (\eg changing
hydrophobicity) gradually shift it to the strong-interaction
regime and finally to the polysoap regime; the {\it cac} is
predicted to decrease from a value close to the {\it cmc} to much
lower values and finally to disappear.

In spite of the vast experimental literature available on
polymer--surfactant systems, additional experiments are still
required in order to verify the theory presented in this work. In
particular, measurement of polymer statistics at the {\it cac}
(\ie dependence of size on polymerization degree) may provide a
clear verification of the partial-collapse conjecture.

We have presented a scaling function relating the {\it cac} and
{\it cmc} and demonstrated its universal features. The scaling
function was explicitly calculated in a mean-field approximation
and at the next level beyond mean field.
It is worth noting, however, that we expect the scaling
law of eqs \ref{cac_general} and \ref{cac_blob} to be of more
general validity than any specific model discussed here.
It should be interesting, therefore,
to gain more information on the scaling law, \eg by computer
simulations, and check the analytic results. We have pointed at
interesting similarities between the effect of small associating
molecules on polymer conformation and general results concerning
the effect of impurities on critical phenomena.

One future extension of this work would be to apply the partial
collapse approach to more concentrated solutions, where the onset
of self-assembly involves many-chain effects and leads to
interesting phase behavior and
gelation.\cite{review_Lindman,review_phase}
 Another direction may be to
consider more complicated polymers such as polypeptides, where
surfactant binding was shown to promote the formation of secondary
structures.\cite{helix}  In addition, the partial-collapse
approach is valid only for flexible polymers, as demonstrated in
Section~\ref{sec_blob}. The interaction of stiff polymers with small
associating molecules is governed by different physics,
requiring a separate treatment.\cite{stiff}

\acknowledgments
%----------------------------
We are grateful to R. Zana for many illuminating discussions. We
greatly benefited from conversations and correspondence with T.
Garel, B. Harris, I. Iliopoulos, L. Leibler, B. Lindman, H.
Orland, L. Piculell, Y. Rabin and T. Witten. Partial support from
the Israel Science Foundation founded by the Israel Academy of
Sciences and Humanities --- Centers of Excellence Program,
and the US--Israel Binational Science Foundation (BSF) under
grant No.\ 98-00429,
is gratefully
acknowledged. HD would like to thank the Clore Foundation for
financial support.

\section*{Appendix A: The Scaling Function beyond Mean Field Theory}
%-------------------------------------------------------------------
\renewcommand{\theequation}{A.\arabic{equation}}
\setcounter{equation}{0}

We present in detail a statistical-mechanical
model leading to explicit expressions for the scaling function,
$F(x)$, introduced in Section~\ref{sec_cac}.
A systematic expansion is derived, which yields
the simple mean-field result, eq~\ref{cac}, as a leading
order, yet allowing us to proceed beyond the mean-field
approximation.

Consider $P$ polymer chains of
$N$ monomers each, which are immersed in a dilute solution
containing $S$ small molecules (\eg surfactants). We use the
grand-canonical ensemble, where $S$ is not fixed but controlled by
a chemical potential, $\mu$. The coordinates of the monomers are
denoted by $\{\vecx^p_n\}_{p=1\ldots P,n=1\ldots N}$ and those of
the small solute molecules are $\{\vecy_s\}_{s=1\ldots S}$. Let
the potentials of solute-solute, monomer-monomer, and
monomer-solute interactions be, respectively, $U(\vecr-\vecr')$,
$V(\vecr-\vecr')$, and $W(\vecr-\vecr')$.
The partition function of the system is
\begin{eqnarray}
  Z &=& \frac{1}{P!} \sum_{S=0}^\infty \frac{1}{S!}\rme^{\mu S}
  \int\prod_{p=1}^P\prod_{n=1}^N\rmd\vecx^p_n
  \prod_{s=1}^S\rmd\vecy_s
  \exp(-\cH_{\rm id} - \cH_{\rm int})
  \nonumber\\
  \cH_{\rm int} &=&
  \half\sum_{p\neq p'}\sum_{n\neq n'} V(\vecx^p_n-\vecx^{p'}_{n'})
  + \half\sum_{s\neq s'} U(\vecy_s - \vecy_{s'})
  + \sum_{p}\sum_{n}\sum_{s} W(\vecx^p_n - \vecy_s),
\end{eqnarray}
where all energy and interaction parameters are given in units of
$k_{\rm B}T$, the thermal energy, and
$\cH_{\rm id}\{\vecx^p_n\}$ is the Hamiltonian of $P$
ideal Gaussian chains.\cite{Gaussian}  Our aim is to
trace out the degrees of freedom of the small molecules
($\{\vecy_s\}$) and find the resulting effective interaction
between monomers.

We introduce continuous densities for the two species,
\[
  c(\vecr) \equiv \sum_{p,n}\delta(\vecr-\vecx^p_n),\ \ \
  \varphi(\vecr) \equiv \sum_{s}\delta(\vecr-\vecy_s),
\]
and their conjugate fields,
$\gamma(\vecr)$ and $\psi(\vecr)$, respectively,
such that
\begin{eqnarray}
  \delta\left[c(\vecr) - \sum_{p,n}\delta(\vecr-\vecx^p_n) \right]
  &=& \int\cD\gamma \exp\left\{ -i\gamma\left[ c(\vecr) -
  \sum_{p,n}\delta(\vecr-\vecx^p_n) \right] \right\}
\nonumber\\
  \delta\left[\varphi(\vecr) - \sum_{s}\delta(\vecr-\vecy_s) \right]
  &=& \int\cD\psi \exp\left\{ -i\psi\left[ \varphi(\vecr) -
  \sum_s \delta(\vecr-\vecy_s) \right] \right\}.
\end{eqnarray}
The partition function is then rewritten as
\begin{eqnarray}
  Z &=& \int\cD c\cD\varphi\cD\gamma\cD\psi
  \exp(-\cH_{\rm cont}) \times\zeta_{\rm p}\times\zeta_{\rm s}
\nonumber\\
  \cH_{\rm cont} &=& \int\rmd\vecr\rmd\vecr'\left[
  \half c(\vecr)V(\vecr-\vecr')c(\vecr') +
  \half \varphi(\vecr)U(\vecr-\vecr')\varphi(\vecr') +
  c(\vecr)W(\vecr-\vecr')\varphi(\vecr')\right]
\nonumber\\
  & & - i\int\rmd\vecr[\gamma(\vecr)c(\vecr) +
  \psi(\vecr)\varphi(\vecr)]
\nonumber\\
  \zeta_{\rm p} &=& \frac{1}{P!}\int\prod_{p,n}\rmd\vecx^p_n
  \exp[-\cH_{\rm id} - i\sum_{p,n}\gamma(\vecx^p_n)]
\nonumber\\
  \zeta_{\rm s} &=& \sum_{S=0}^\infty \frac{1}{S!}\rme^{\mu S}
  \int\prod_{s=1}^S\rmd\vecy_s
  \exp[-i\sum_{s=1}^S\psi(\vecy_s)]
  = \exp[\phib\int\rmd\vecr\;\rme^{-i\psi(\vecr)}].
\end{eqnarray}
In the last equation we have exploited the independence of the
integral term on $s$ and the expansion of the exponential function
in power series,
where $\phib\equiv\rme^\mu$ is the average solute concentration in
the bulk reservoir. (We assume an ideal solution of small solute
molecules in the bulk
reservoir, \ie a vanishing $\psi$ and $\mu=\log\varphi_{\rm b}$.)

It is convenient to transform to Fourier space,
$\tilde{f}_k \equiv \int\rmd\vecr\;\rme^{-i\veck\cdot\vecr} f(\vecr)$,
whereupon $\cH_{\rm cont}$ becomes
\[
  \cH_{\rm cont} = \int\rmd\veck\left[
  \half \tV_k|\tc_k|^2 + \half \tU_k |\tphi_k|^2
  + (\tW_k \tc_k+i\tpsi_k)\tphi_{k}
  + i\tgamma_k \tc_{k} \right].
\]
Tracing over the solute concentration profile, $\varphi_k$,
is straightforward, giving (up to a constant factor)
\begin{equation}
  Z = \int\cD \tc_k\cD\tgamma_k\cD\tpsi_k
  \exp\left\{ \int\rmd\veck \left[ \half \tU_k^{-1}
  (\tW_k \tc_k+i\tpsi_k)^2
  - \half \tV_k|\tc_k|^2 - i\tgamma_k \tc_{k} \right]\right\}
  \times\zeta_{\rm p}\times\zeta_{\rm s}.
\label{Z_after_phi}
\end{equation}
In the usual case, where the potentials of interaction,
$U(\vecr,\vecr')$, $V(\vecr,\vecr')$ and $W(\vecr,\vecr')$,
depend only on $(\vecr-\vecr')$,
they are diagonal in $k$-space and can be simply inverted,
\eg
$\tU_k^{-1}=1/\tU_k$.

In order to trace out also the solute field, $\psi$, we proceed by
an expansion of $\zeta_{\rm s}$ in small $\psi$. Physically,
$\psi$ accounts for interactions between the small solute
molecules. The small parameter of the expansion, therefore, is the
strength of solute-solute correlations in the solution.
The following calculation is expected to give good results in the
regime of strong polymer--solute interaction [$F(x\ll 1)$],
and less accurate results
in the limit of weak polymer--solute interaction
[$F(x\gg 1)$], where solute--solute
correlations become important.

\subsection*{Gaussian Approximation}
%-----------------------------------

In the Gaussian approximation $\zeta_{\rm s}$ is expanded to 2nd
order in $\psi$,
\begin{equation}
  \zeta_{\rm s} \simeq \mbox{const}\times
  \exp\left[ -\phib\int\rmd\vecr (i\psi + \half\psi^2) \right]
  = \mbox{const}\times
  \exp\left[-\phib( \half\int\rmd\veck |\tpsi_k|^2 - i\tpsi_{0}) \right],
\end{equation}
where $\tpsi_0\equiv\tpsi_{k=0}=\int\rmd\vecr\psi(\vecr)$.
Substituting this expression in eq~\ref{Z_after_phi} we get
\begin{eqnarray}
  Z &\simeq& Z_2 = \int\cD \tc_k\cD\tgamma_k\cD\tpsi_k
  \exp(-\cH_2) \times\zeta_{\rm p}
 \nonumber\\
  \cH_2 &=& \int\rmd\veck \left[
  \half(\tU_k^{-1}+\phib)|\tpsi_k|^2 - i\tW_k \tU_k^{-1}\tc_k\tpsi_k
  +\half(\tV_k-\tW_k^2\tU_k^{-1})|\tc_k|^2 + i\tgamma_k \tc_k \right]
\nonumber\\
  & & - i\phib\tpsi_0.
\label{Z_before_psi}
\end{eqnarray}
For $c=0$, instability with respect to small perturbations in
$\psi$ will occur if there exists $\veck$ such that the
coefficient of $|\psi_k|^2$ vanishes, \ie $1/U_k+\phib=0$. The
{\it cmc} is therefore identified as
\begin{equation}
  \cmc = \min_\veck (-1/\tU_k).
\end{equation}
Tracing $\tpsi_k$ out of eq~\ref{Z_before_psi} gives (again, up
to a constant factor)
\begin{eqnarray}
 Z_2 &=& \int\cD\tc_k\cD\tgamma_k \exp \left\{ \int\rmd\veck \left[
  -\half(\tV_k - \frac{\phib \tW_k^2}{1+\phib\tU_k} )|\tc_k|^2
  -\frac{\phib(\tW_k\tc_k+\phib\tU_k/2)}{1+\phib\tU_k} \delta(\veck)
  - i\tgamma_k \tc_k \right] \right\} \nonumber\\
  & & \times\zeta_{\rm p}.
\label{Z_after_psi}
\end{eqnarray}
Thus, as we found in Section~\ref{sec_cac},
the small solute induces an effective reduction in the
potential between monomers, which becomes more significant
as the {\it cmc} is approached,
\begin{equation}
  \tV_{k,{\rm eff}} = \tV_k - \frac{\phib\tW_k^2}
  {1+\phib\tU_k},
\label{Vkeff}
\end{equation}
The second term in eq~\ref{Vkeff} can be identified as the
Fourier transform of the induced potential between monomers, as
is phenomenologically introduced in Appendix~B,
$\tilde{\Phi}_k=-\phib\tW_k^2/(1+\phib\tU_k)$.

As in the previous sections, the {\it cac} is assumed to
correspond to the vanishing of the effective interaction,
\begin{equation}
  \cac = \min_\veck \frac{-1/\tU_k}
  {1-\tW_k^2/(\tV_k \tU_k)}.
\end{equation}
If we neglect the finite range of the various interactions and
substitute the corresponding simplified potentials (taking the
monomer-monomer interaction as repulsive and the monomer-solute
and solute-solute ones as attractive, \ie $u,v,w>0$),
\[
  V(\vecr-\vecr')=v\delta(\vecr-\vecr'),\ \
  U(\vecr-\vecr')=-u\delta(\vecr-\vecr'),\ \
  W(\vecr-\vecr')=-w\delta(\vecr-\vecr')
\]
our mean-field result \ref{cac} is recovered,
\begin{eqnarray}
  \cac &=& \cmc F[v/(w^2\cmc)]
 \nonumber\\
  F(x) &=& 1/(1+1/x).
\end{eqnarray}

\subsection*{Beyond Gaussian Approximation}
%-----------------------------------------

We now calculate the first correction to the Gaussian
approximation (\ie mean field) considering terms of 3rd order in
$\psi$,
\begin{eqnarray}
  Z &\simeq& Z_3 = \int\cD c\cD\gamma\cD\psi
  \exp( -\cH_2 - \cH_3 ) \times\zeta_{\rm p}
 \nonumber\\
  \cH_3 &=& -\frac{i}{6}\phib \int\rmd\vecr\;\psi^3.
\label{Z3}
\end{eqnarray}
To the same order of approximation we can write
\begin{equation}
  \int\rmd\vecr\;\psi^3 \simeq \int\rmd\vecr\langle\psi^3\rangle
  = \int\rmd\veck\rmd\veck' \langle \tpsi_k\tpsi_{k-k'}
  \tpsi_{k'} \rangle,
\label{Z3k}
\end{equation}
where $\langle\cdots\rangle$ denotes a thermal average using the
Gaussian approximation ($\cH_2$). By means of our results for
$Z_2$, eqs \ref{Z_before_psi} and \ref{Z_after_psi}, we find
\begin{eqnarray}
  g_1(\veck) &\equiv& \langle\tpsi_k\rangle =
  i\frac{\tW_k\tc_k + \phib\tU_k\delta(\veck)}
  {1+\phib\tU_k}
 \nonumber\\
  g_2(\veck,\veck') &\equiv& \langle\tpsi_k\tpsi_{k'}\rangle
  = g_1(\veck)g_1(\veck') + \frac{\tU_k\delta(\veck-\veck')}
  {1+\phib\tU_k}
 \nonumber\\
  g_3(\veck,\veck',\veck'') &\equiv& \langle\tpsi_k
  \tpsi_{k'}\tpsi_{k''}\rangle
  = 3g_2(\veck,\veck')g_1(\veck'') - 2g_1(\veck)g_1(\veck')
  g_1(\veck'') \nonumber\\
  &=& g_1(\veck)g_1(\veck')g_1(\veck'') +
  \frac{3\tU_k\delta(\veck-\veck')} {1+\phib\tU_k}
  g_1(\veck'').
\end{eqnarray}
The expression for $g_3$ should now be integrated according to
eq~\ref{Z3k}. However, focusing on the effective pairwise
potential between monomers we look for terms which are quadratic
in $c$. There is only one such term, coming from the integration
of $g_1(\veck)g_1(\veck-\veck')g_1(\veck')$. This gives
\begin{equation}
  \cH_3 = -\frac{\phib^2\tU_0} {2(1+\phib\tU_0)}
  \int\rmd\veck \frac{\tW_k^2}{(1+\phib\tU_k)^2} |\tc_k|^2 \ + \
  \mbox{non-quadratic terms}.
\end{equation}
The effective potential, therefore, is
\begin{equation}
  \tV_{k,{\rm eff}} = \tV_k -
  \frac{\phib \tW_k^2}{1+\phib \tU_k} \left[
  1 + \frac{\phib \tU_0} {(1+\phib \tU_0)(1+\phib \tU_k)}
  \right],
\end{equation}
where the second term can be identified, again, as the induced
potential, $\tilde{\Phi}_k$.
Substituting the simpler potentials, $\tV_k=v$, $\tU_k=-u$,
$\tW_k=-w$, we find the corrected scaling function (given in an
implicit form),
\begin{eqnarray}
  \cac &=& \cmc F[v/(w^2\cmc)]
\nonumber\\
  F^{-1}(y) &=& \frac{y(1-y+y^2)}{(1-y)^3}
\nonumber\\
  F(x) &\simeq& \left\{
  \begin{array}{ll}
  x - 2x^2 & \ \ \ x\ll 1 \\
  1 - 1/x^{1/3} & \ \ \ x\gg 1
  \end{array}
  \right.
\end{eqnarray}

\section*{Appendix B: Scaling Analysis of Partial Collapse}
%----------------------------------------------------------
\renewcommand{\theequation}{B.\arabic{equation}}
\setcounter{equation}{0}

Based on a scaling analysis we obtain a more detailed description
of the polymer at partial collapse, leading to more accurate
predictions regarding the {\it cac}.
(The reasoning presented in this Appendix is similar to
that of ref~\cite{deGennes}.)
As shown in Section~\ref{sec_cac}, the interaction with the
small molecules induces attractive correlations between monomers
in the chain.
In the following analysis, therefore, the small molecules
(\eg surfactants) do not appear explicitly, but are
represented by an effective
attractive potential exerted between monomers,
\[
  \Phi(r) = -e^2\chi(r/\xi).
\]
Following the notation of ref~\cite{deGennes}, $e^2$ is a
coupling constant, $\xi$ a correlation length, and $\chi(r/\xi)$ a
dimensionless function which decays fast to zero for $r>\xi$
The two microscopic
parameters, $e^2$ and $\xi$, are to be related to our
phenomenological interaction parameter, $w$. Assuming weak
correlations, $\Phi<1$ (in units of $\kB T$), we readily obtain
for the effective excluded-volume parameter of the chain,
\[
  v_{\rm eff} = v + \int\rmd\vecr\; \Phi(r) = v - k_1 e^2 \xi^3,
\]
where $k_1$ is a dimensionless constant. Comparing to
eq~\ref{veff} we can identify
\begin{equation}
  e^2 \sim \frac{v_{\rm ps}} {\xi^3}.
\label{coupling}
\end{equation}

In accordance with the model presented in Section \ref{sec_blob},
we consider a chain of blobs of size $\xi$, each
containing $g$ statistical segments (see
Figure~\ref{fig_blob}). The potential of interaction between
two blobs consists of a hard-core part for $r<\xi$, and an
attractive part for $r>\xi$ coming from the integrated interaction
of $g^2$ pairs of monomers,
\[
  \Phi_{\rm blob}(r) \sim \left\{
  \begin{array}{ll}
  \infty  & \ \ \ r<\xi \\
  g^2 \Phi(r)  & \ \ \ r>\xi
  \end{array}
  \right.
\]
The resulting excluded-volume parameter for the blobs
is
\begin{equation}
  v_{\rm blob} = \int\rmd\vecr [1 - \rme^{-\Phi_{\rm blob}(r)}]
  \simeq k_2\xi^3 - k_3\xi^3\rme^{k_4 g^2 e^2},
\end{equation}
where $k_2$,$k_3$,$k_4$ are dimensionless constants.
(Note that although $\Phi<1$, $\Phi_{\rm blob}\sim g^2\Phi$
may be large.)
The condition for partial collapse is $v_{\rm blob}=0$, \ie
\begin{equation}
  g^2 e^2 = \log(k_2/k_3)/k_4 = \mbox{const}.
\label{blobcollapse}
\end{equation}
Two relations for $e^2$, $g$ and $\xi$ have been obtained
(eqs \ref{coupling} and \ref{blobcollapse}).
A third relation comes from the statistics of the
polymer, \ie the power law relating
the blob size and number of segments in the blob,
\begin{equation}
  \xi \sim g^\nu a^{z} v^{(1-z)/3},
\label{statistics}
\end{equation}
where $a$ is the length of a Kuhn segment.
%(In the case of
%excluded-volume repulsion in three dimensions, the Flory argument
%gives $\nu=3/5$ and $z=2/5$.)
From the three relations ---
\ref{coupling}, \ref{blobcollapse} and \ref{statistics} ---
we get
\begin{eqnarray}
  g &\sim& \left(\frac{v}{v_{\rm ps}}\right)^{1/\alpha}
    \left(\frac{a^3}{v}\right)^{z/\alpha}
 \nonumber\\
  \xi &\sim& \left(\frac{v}{v_{\rm ps}}\right)^{\nu/\alpha}
    \left(\frac{a^3}{v}\right)^{\nu z/\alpha}
    a^z v^{(1-z)/3}
 \nonumber\\
  \alpha &\equiv& 2-3\nu.
\end{eqnarray}
%

%----------------------------------------------------
% References
%----------------------------------------------------

%---------------------------------------------------
%\end{multicols}

\newpage

%----------------------------------------------
% Tables
%----------------------------------------------

\begin{table} %dependence on I
\caption[]{Dependence of {\it cac} on polyelectrolyte ionization
degree in the strong-interaction regime \cite{c_half}}
\begin{tabular}{cccdddc}
\setlength{\baselineskip}{0.5\baselineskip}
  polymer & surfactant & salt & $c_{\rm salt}$ (mM) & $I$ & $\cac$ (mM) &
reference \\
\tableline
   PMAMVE
   %\tablenote{Poly(maleic acid-co-methyl vinyl ether),
  %dodecyltrimethylammonium bromide, KBr, ref~\cite{Benrraou92}.}
  & DTAB & KBr & 5  & 0.5 & 0.16 & \cite{Benrraou92} \\
        &   &   &    & 1   & 0.16 & \\
\tableline
   PMAEVE
   %\tablenote{Poly(maleic acid-co-ethyl vinyl ether),
  %dodecyltrimethylammonium bromide, KBr, ref~\cite{Benrraou92}.}
  & DTAB & KBr & 5  & 0.5 & 0.09 & \cite{Benrraou92} \\
        &   &   &    & 1   & 0.1 & \\
\tableline
   PMAMVE
   %\tablenote{Poly(maleic acid-co-methyl vinyl ether),
  %dodecyltrimethylammonium chloride, NaCl, ref~\cite{Anthony96}.}
  & DTAC & NaCl & 20 & 0   & 6 & \cite{Anthony96} \\
        &   &   &    & 1   & 5 & \\
\tableline
    chitosan & SDS & NaBr & 20 & 0.76 & 0.028 & \cite{Wei93} \\
        &   &   &   & 0.84 & 0.028 & \\
        &   &   &   & 0.99 & 0.028 & \\
\tableline
   PMASt
   %\tablenote{Poly(maleic acid-co-styrene),
  %dodecylpyridinium chloride, NaCl, ref~\cite{Shimizu86}.}
  & C$_{12}$PyCl & NaCl & 25 & 0.5 & 0.025 & \cite{Shimizu86} \\
        &   &    &    & 1   & 0.025 & \\
\tableline
   PMAIn
   %\tablenote{Poly(maleic acid-co-indene),
  %dodecylpyridinium chloride, NaCl, ref~\cite{Shimizu86}.}
  & C$_{12}$PyCl & NaCl & 25 & 0.5 & 0.012 & \cite{Shimizu86} \\
         &   &   &    & 1   & 0.016 & \\
\tableline
    PMAEVE & C$_{12}$PyCl & NaCl & 25 & 0.5 & 0.15 & \cite{Shimizu86} \\
        &   &  &  & 1 & 0.22 & \\
\tableline
    PMAE & C$_{12}$PyCl & NaCl & 25 & 0.5 & 0.13 & \cite{Shimizu86} \\
        &   &  &  & 1 & 0.27 & \\
\tableline
    PAA & TTAB & NaBr & 10 & 0.14 & 0.0068 & \cite{Kiefer93} \\
        &   &   &   & 0.26 & 0.0092 & \\
        &   &   &   & 0.5 & 0.025 & \\
        &   &   &   & 1 & 0.029 & \\
\tableline
    PAA & DTAB & NaCl & 30 &
    pH=5.3\tablenote{pH
values were reported; $I$ depends monotonously on pH.}
    & 0.4 & \cite{Chandar88} \\
        &   &   &   & 6.4 & 0.7 & \\
        &   &   &   & 8.1 & 0.9 & \\
        &   &   &   & 10.8 & 0.95 & \\
\tableline
    PVS & DTAB & KCl & 10 & 0.18 & 0.18 & \cite{Satake90} \\
        &   &   &   & 0.34 & 0.11 & \\
        &   &   &   & 0.50 & 0.12 & \\
        &   &   &   & 0.68 & 0.06 & \\
        &   &   &   & 0.74 & 0.05 & \\
        &   &   &   & 1 & 0.05 & \\
\setlength{\baselineskip}{2\baselineskip}
\end{tabular}
\label{tab_I}
\end{table}

\begin{table} %cac/cmc in weak regime
\caption[]{Relation between {\it cac} and {\it cmc} in the
weak-interaction regime}
\begin{tabular}{cccdddc}
  polymer & surfactant & salt/counterion & $\cac$ (mM) & $\cmc$ (mM)
  & $\cac/\cmc$ & reference \\
\tableline
   PEO
   %\tablenote{Poly(ethylene oxide), dodecylsulfate,
   %ref~\cite{Wang98}.}
  & DS$^-$ & NaDS & 4.2 & 8.2 & 0.51 & \cite{Wang98} \\
        &  & LiDS & 3.9 & 7.7 & 0.51 & \\
        &  & NaDS, 0.1M NaCl & 0.8 & 1.4 & 0.57 & \\
\tableline
   PEO & SDS & no salt & not & 8.2 & 0.65 & \cite{Cabane82} \\
  &     & 0.075M NaCl & reported\tablenote{
   %Poly(ethylene oxide), sodium dodecylsulfate,
   %ref~\cite{Cabane82}.
   Only the ratio $\cac/\cmc$ was reported.
   {\it cmc} data taken from refs~\cite{Corti81,Kamenka97}.}
  &   & 0.67 & \\
  &     & 0.15M NaCl &     &   & 0.69 & \\
  &     & 0.2M NaCl &      & 0.94  & 0.76 & \\
  &     & 0.4M NaCl &      & 0.59  & 0.78 & \\
\tableline
   PVP
   %\tablenote{Poly(vinylpyrrolidone), sodium dodecylsulfate,
   %ref~\cite{Murata73}.}
  & SDS & 0.01M NaCl & 2.2 & 5.5 & 0.40 & \cite{Murata73} \\
  &     & 0.1M NaCl  & 0.84 & 1.9 & 0.44 & \\
\tableline
   PAA
   %\tablenote{Poly(acrylic acid), C$_n$E$_8$,
   %ref~\cite{Anghel98}.}
  & C$_{10}$E$_8$ & --- & 0.7 & 1 & 0.7 & \cite{Anghel98} \\
  & C$_{12}$E$_8$ &  & 0.063 & 0.08 & 0.79 & \\
  & C$_{14}$E$_8$ &  & 0.007 & 0.009 & 0.78 & \\
\end{tabular}
\label{tab_weak}
\end{table}

\section*{Figure Captions}
%-------------------------

\begin{figure}[tbh]
\vspace{3cm} \centerline{ \epsfysize=8cm
\hbox{\epsffile{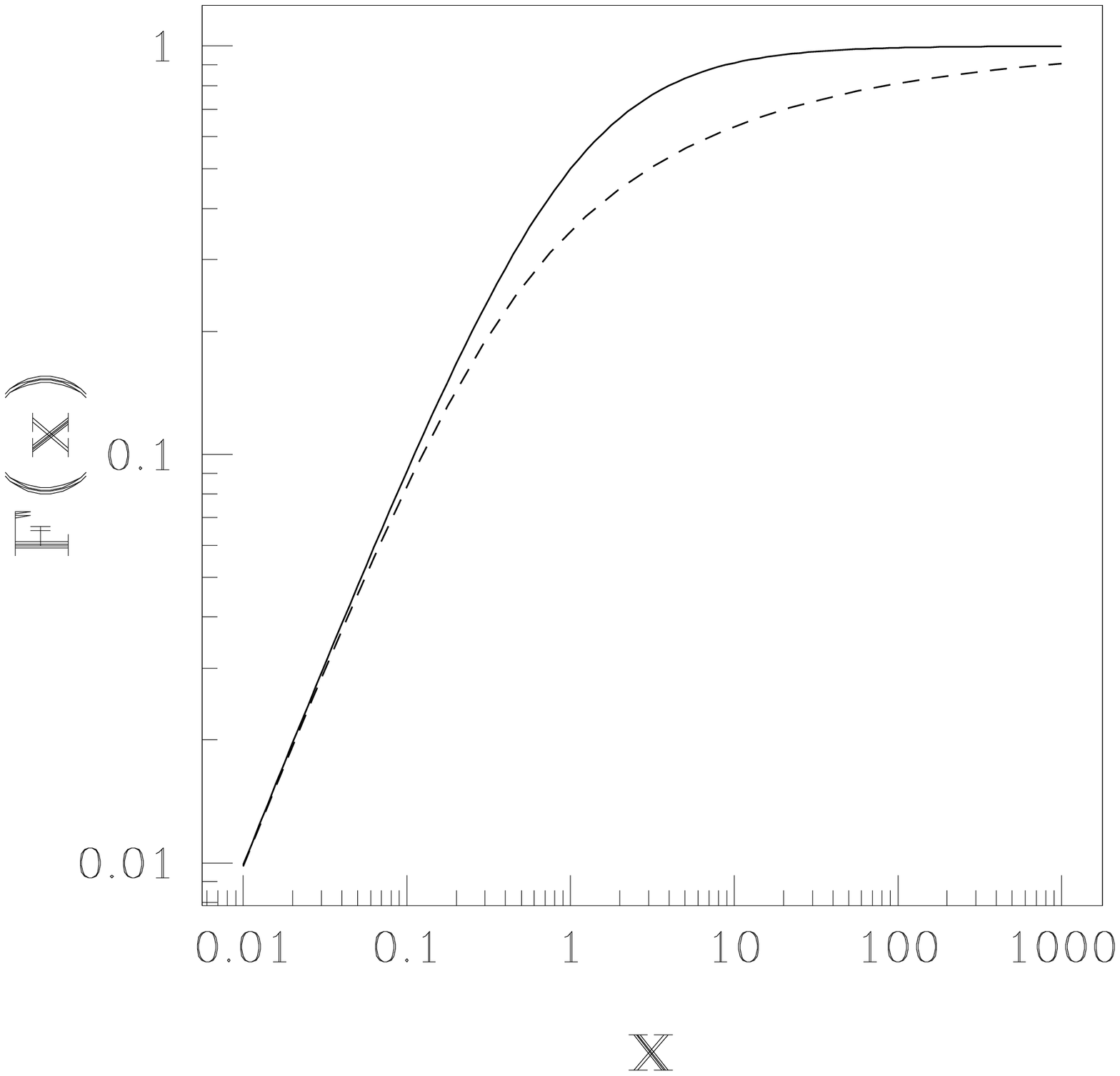}}}
\caption[]{}%
\label{fig_F}
\end{figure}

\begin{figure}[tbh]
%\vspace{2cm}
\centerline{ \epsfysize=8cm
\hbox{\epsffile{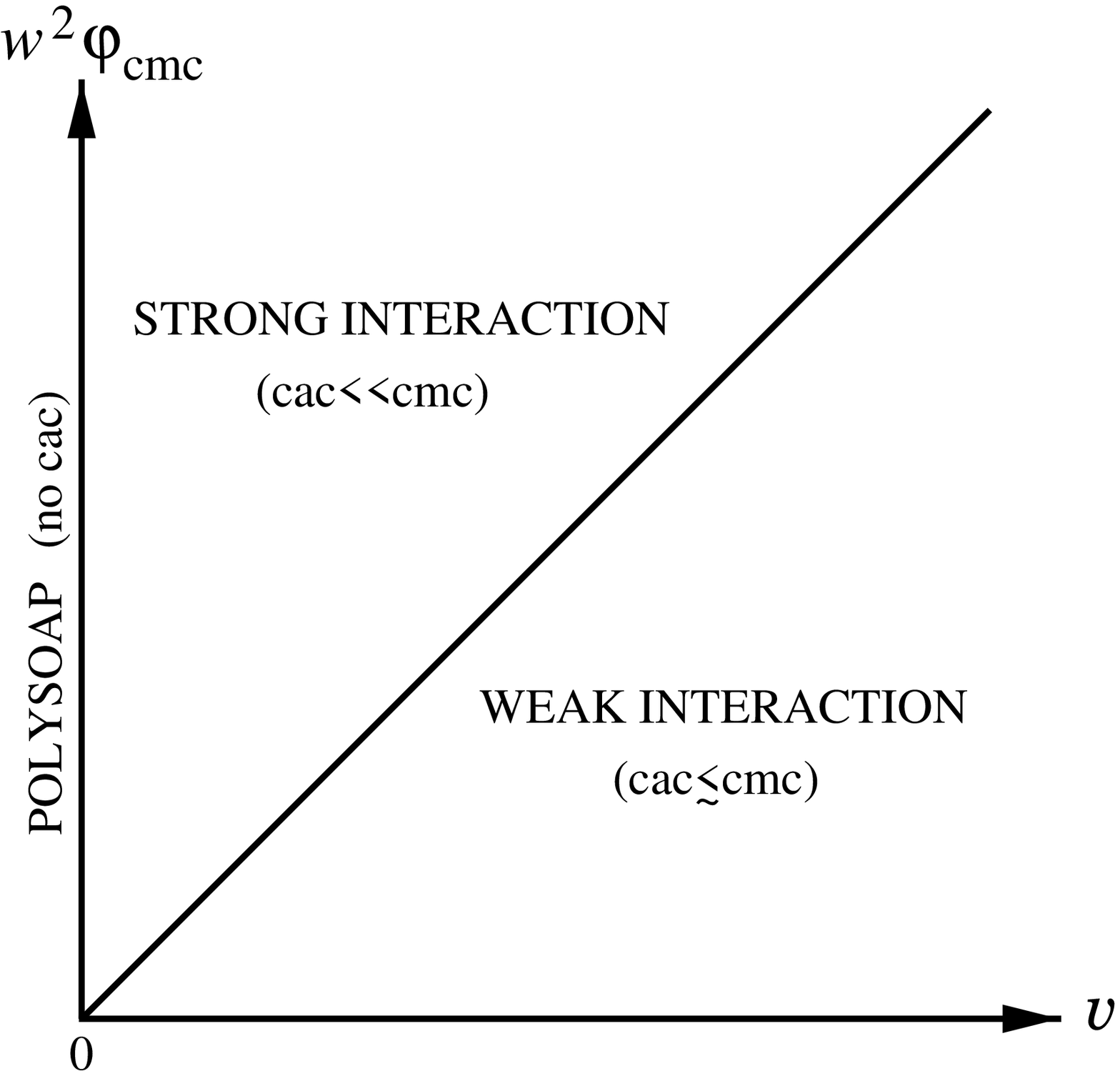}}}
\caption[]{}%
\label{fig_diag}
\end{figure}

\begin{figure}[tbh]
%\vspace{2cm}
\centerline{ \epsfysize=8cm
\hbox{\epsffile{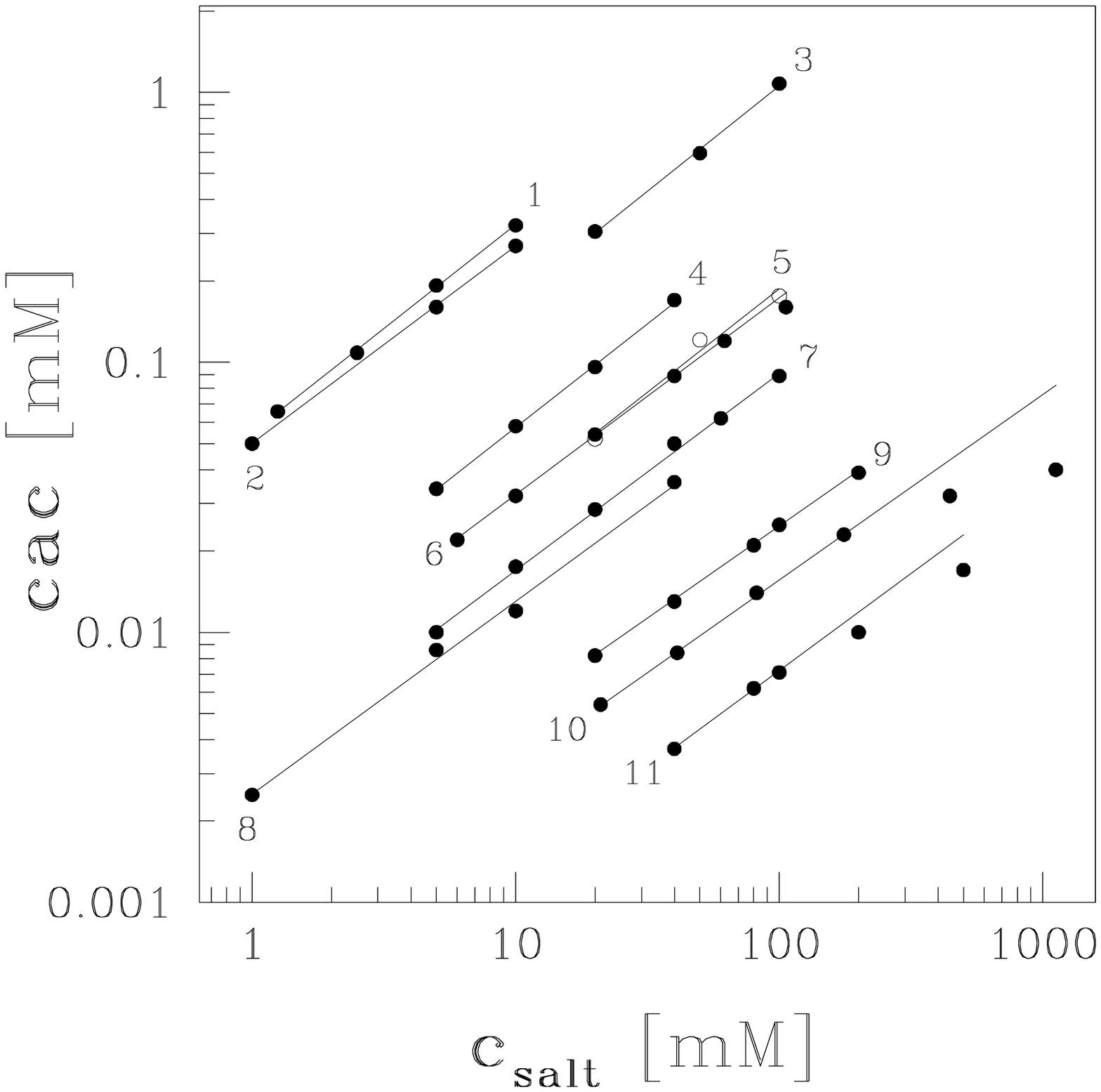}}}
\caption[]{} %
\label{fig_salt}
\end{figure}

\begin{figure}[tbh]
\vspace{3cm} \centerline{ \epsfysize=8cm
\hbox{\epsffile{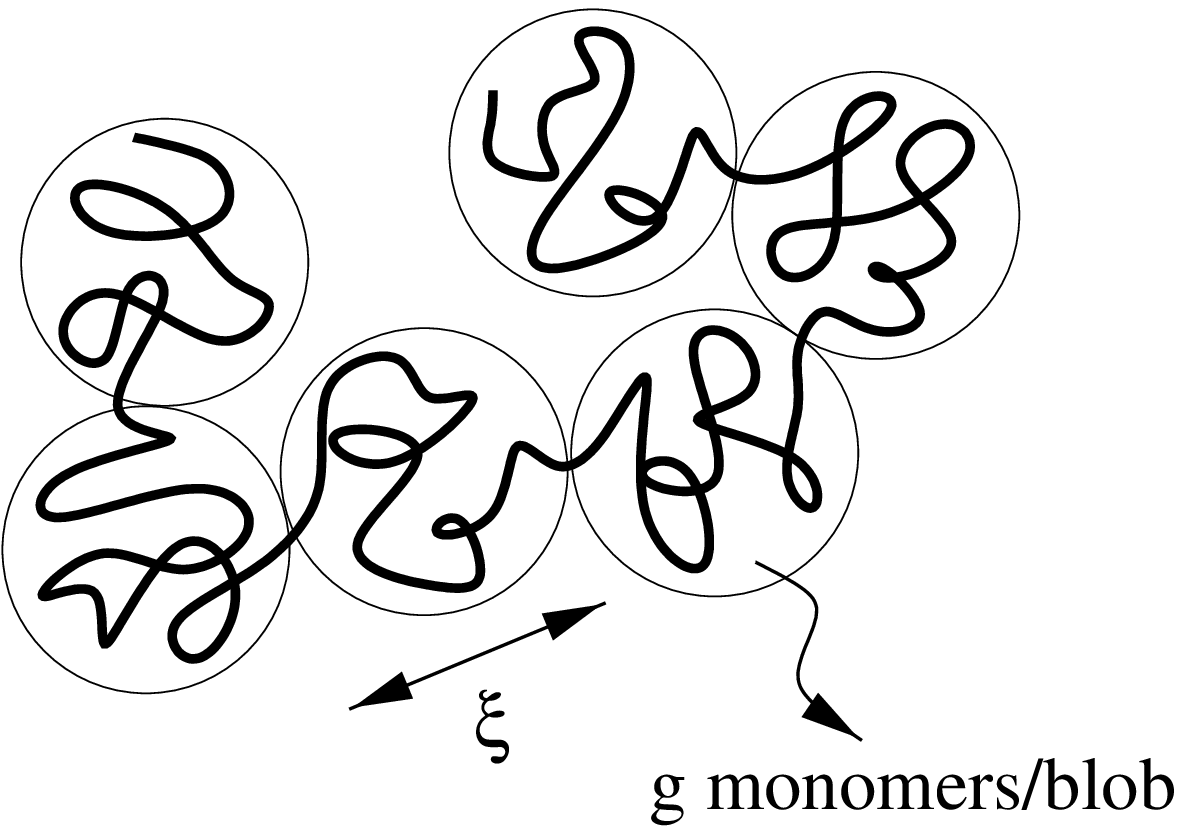}}}
\caption[]{} %
\label{fig_blob}
\end{figure}

\begin{figure}[tbh]
%\vspace{2cm}
\centerline{ \epsfysize=8cm
\hbox{\epsffile{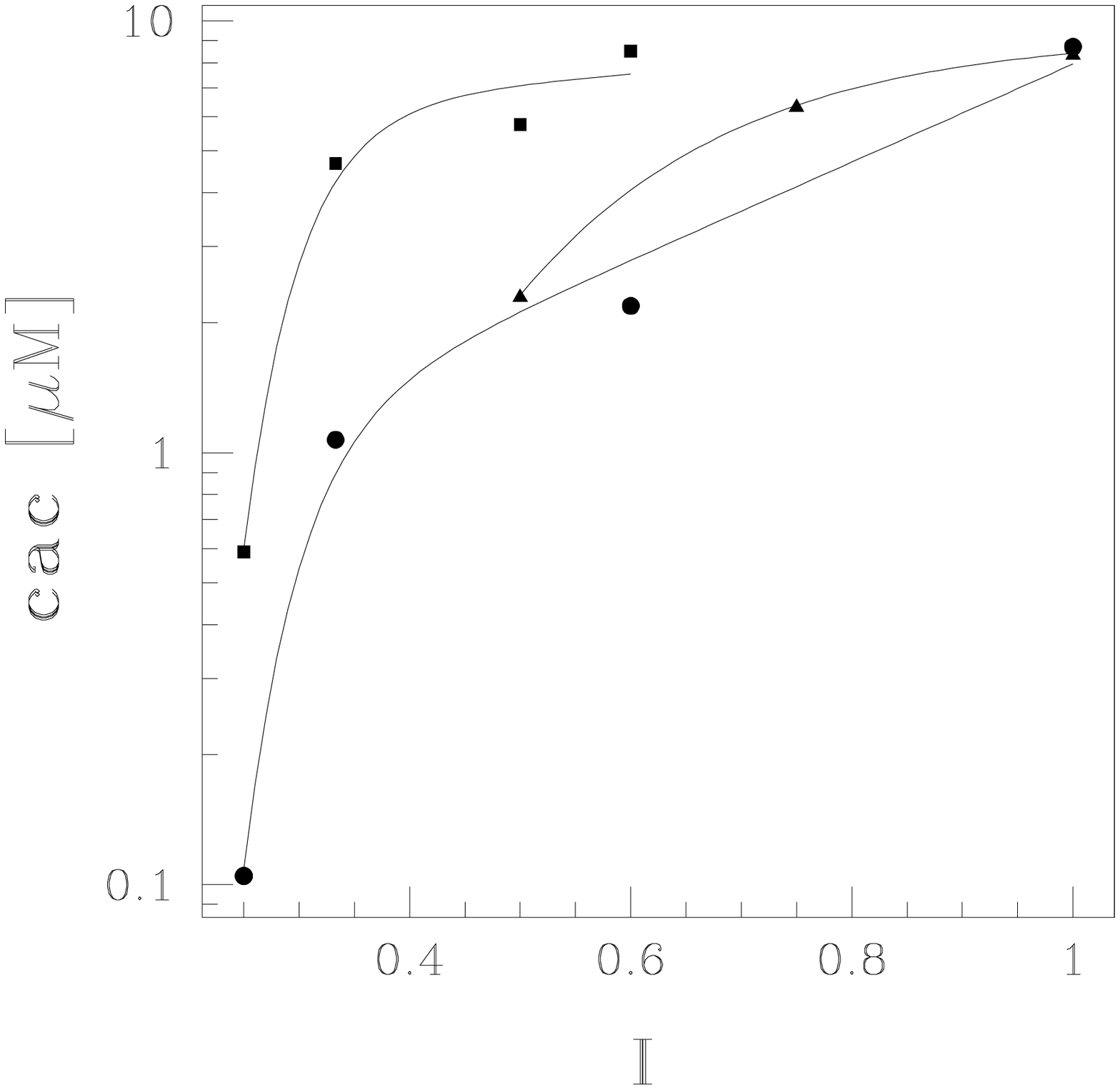}}}
\caption[]{}%
\label{fig_amphi}
\end{figure}

\begin{itemize}

\item[{\bf Figure~\ref{fig_F}}]
The scaling function $F(x)$. Solid line --- mean-field calculation
(eq~\ref{cac}); dashed --- beyond mean field
(eq~\ref{F_beyond}).

\item[{\bf Figure~\ref{fig_diag}}]
Summary of self-assembly regimes: (i) a strong-interaction regime
($w^2\cmc>v$) where $\cac\ll\cmc$, corresponding experimentally to
systems containing oppositely charged species; (ii) a
weak-interaction regime ($w^2\cmc<v$) where $\cac\lesssim\cmc$,
corresponding to systems where at least one of the species is
neutral; (iii) a polysoap regime ($v=0$) where association is
gradual (no {\it cac}), corresponding to polymers which form
intra-chain aggregates by themselves.

\item[{\bf Figure~\ref{fig_salt}}]
Dependence of {\it cac} on monovalent salt concentration in various
polyelectrolyte--oppositely charged surfactant
systems.\cite{c_half}  From top to bottom: (1) NaPA, DTAB, NaBr
(ref~\cite{Hansson95});
(2) %poly(maleic acid-co-methylvinylether)
PMAMVE, DTAB, KBr (ref~\cite{Benrraou92}); (3) PVS, C$_{10}$PyBr,
NaBr (ref~\cite{Shirahama84}); (4) NaDxS, C$_{11}$PyBr, NaCl
(ref~\cite{Malovikova84}); (5) PAS, SDeS, NaCl
(ref~\cite{Shirahama81});
(6) % sodium dextran sulfate (SDxS),
NaDxS, %dodecyltrimethylammonium bromide
DTAB, NaCl (ref~\cite{Hayakawa82}); (7) NaDxS, C$_{12}$PyCl, NaCl
(ref~\cite{Malovikova84}); (8) PMABVE, DTAB, KBr
(ref~\cite{Benrraou92}); (9) NaDxS, C$_{13}$PyBr, NaCl
(ref~\cite{Malovikova84});
(10) %sodium polystyrenesulfunate
PSS, DTAB, NaCl (ref~\cite{Hayakawa82}); (11) NaDxS,
C$_{14}$PyBr, NaCl (ref~\cite{Malovikova84}). Fitted slopes lie
in the range 0.68--0.77.

\item[{\bf Figure~\ref{fig_blob}}]
Schematic sketch of a partially collapsed chain.

\item[{\bf Figure~\ref{fig_amphi}}]
Dependence of {\it cac} on polymer charge close to the polysoap
regime. Triangles --
%poly(maleic acid-co-butylvinylether)
PMABVE, DTAB, 5 mM KBr (ref~\cite{Benrraou92}). For $I<0.5$ this
polymer becomes a polysoap. Circles --
(CH$_2$)$_x$(CH$_2$)$_y$-ionine bromide, SDS, no salt; squares --
with 20 mM NaBr (ref~\cite{Chen98}). A distance of 3 hydrocarbon
groups between charged groups along the backbone has been defined
as $I=1$. The lines are merely guides to the eye.

\end{itemize}

%------------------------------------------------------------
% Figures
%------------------------------------------------------------
\pagebreak

%--------------------------------------------------------------


\begin{thebibliography}{99}

\bibitem{review_general1}
{\it Interactions of Surfactants with Polymers and Proteins};
Goddard, E. D.; Ananthapadmanabhan, K. P., Eds.; CRC: Boca Raton,
FL, 1993.

\bibitem{review_general2}
{\it Polymer-Surfactant Systems}; Kwak, J. C. T., Ed.; Marcel
Dekker: New York, 1998.

\bibitem{review_application}
Goddard, E. D. In ref~\cite{review_general1}, Chapter 10.
Goddard, E. D.; Ananthapadmanabhan, K. P. In
ref~\cite{review_general2}, Chapter 2.

\bibitem{letter}
Diamant, H.; Andelman D.
{\it Europhys. Lett.} {\bf 1999}, {\it 48}, 170.

\bibitem{review_pe_ionic}
Goddard, E. D. In ref~\cite{review_general1}, Chapter 4II.

\bibitem{review_neutral_ionic}
Goddard, E. D. In ref~\cite{review_general1}, Chapter 4I.

\bibitem{review_Anghel}
Saito, S.; Anghel, D. F. In ref~\cite{review_general2}, Chapter
9.

\bibitem{Vasilescu97}
Vasilescu, M.; Anghel, D. F.; Almgren, M.; Hansson, P.; Saito, S.
{\it Langmuir} {\bf 1997}, {\it 13}, 6951.

\bibitem{Anghel98}
Anghel, D. F.; Saito, S.; B\~aran, A.; Iovescu, A. {\it Langmuir}
{\bf 1998}, {\it 14}, 5342.

\bibitem{review_Lindman}
Lindman, B.; Thalberg, K. In ref~\cite{review_general1}, Chapter
5.

\bibitem{Feitosa96}
Feitosa, E.; Brown, W.; Hansson, P. {\it Macromolecules} {\bf
1996}, {\it 29}, 2169. Feitosa, E.; Brown, W.; Vasilescu, M.;
Swanson-Vethamuthu, M. {\it Macromolecules} {\bf 1996}, {\it 29},
6837. Feitosa, E.; Brown, W.; Swanson-Vethamuthu, M. {\it
Langmuir} {\bf 1996} {\it 12}, 5985.

\bibitem{micelle}
Israelachvili, J. {\it Intermolecular and Surface Forces}, 2nd
Ed.; Academic Press: London, 1991, Chapter 17.

\bibitem{review_theory}
Linse, P.; Piculell, L.; Hansson, P. In
ref~\cite{review_general2}, Chapter 5.

\bibitem{micelle_models}
Smith, M. L.; Muller, N. {\it J. Colloid Interface Sci.} {\bf
1975}, {\it 52}, 507.
%Gilanyi, T.; Wolfram, E. {\it Colloid Surf.}
%{\bf 1981}, {\it 3}, 181.
Nagarajan, R. {\it Colloid Surf.} {\bf 1985}, {\it 13}, 1; {\it J.
Chem. Phys.} {\bf 1989}, {\it 90}, 1980. Hall, D. G. {\it J. Chem.
Soc. Faraday Trans. 1} {\bf 1985}, {\it 81}, 885. Ruckenstein, E.;
Huber, G.; Hoffmann, H. {\it Langmuir} {\bf 1987}, {\it 3}, 382.

\bibitem{Shirahama81}
Shirahama, K.; Yuasa, H.; Sugimoto, S. {\it Bull. Chem. Soc. Jpn.}
{\bf 1981}, {\it 54}, 375.

\bibitem{Shirahama84}
Shirahama, K.; Tashiro, M. {\it Bull. Chem. Soc. Jpn.} {\bf 1984},
{\it 57}, 377.

\bibitem{ZimmBragg}
Zimm, B. H.; Bragg, J. K. {\it J. Chem. Phys.} {\bf 1959}, {\it
31}, 526.

\bibitem{polypeptide}
Satake, I.; Yang, J. T. {\it Biopolym.} {\bf 1976}, {\it 15},
2263.

\bibitem{PB_models}
Delville, A. {\it Chem. Phys. Lett.} {\bf 1985}, {\it 118}, 617.
\~Skerjanc, J.; Kogej, K.; Vesnaver, G. {\it J. Phys. Chem.} {\bf
1988}, {\it 92}, 6382.

\bibitem{Levin}
Kuhn, P. S.; Levin, Y.; Barbosa, M. C.
{\it Chem. Phys. Lett.} {\bf 1998}, {\it 298}, 51.

\bibitem{Colby}
Konop, A. J.; Colby, R. H. {\it Langmuir} {\bf 1999}, {\it 15},
58.

\bibitem{Cabane82}
Cabane, B.; Duplessix, R. {\it J. Phys. (Paris)} {\bf 1982},
{\it 43}, 1529.

\bibitem{yes_conformation1}
Abuin, E. B.; Scaiano, J. C. {\it J. Am. Chem. Soc.} {\bf 1984},
{\it 106}, 6274.

\bibitem{yes_conformation2}
Winnik, F. M.; Winnik, M. A.; Tazuke, S. {\it J. Phys. Chem.} {\bf
1987}, {\it 91}, 594. Winnik, F. M.; Ringsdorf, H.; Venzmer, J.
{\it Langmuir} {\bf 1991}, {\it 7}, 912.

\bibitem{yes_conformation3}
Gao, Z.; Wasylishen, R. E.; Kwak, J. C. T {\it J. Phys. Chem.}
{\bf 1991}, {\it 95}, 462.

\bibitem{old_letter}
We have recently demonstrated how the introduction of polymeric
degrees of freedom into such a one-dimensional lattice model leads
to an effective `cooperativity' among bound surfactants. See
Diamant, H.; Andelman, D. {\it unpublished}, cond-mat/9804086.

\bibitem{recent_models}
Wallin, T.; Linse, P. {\it Langmuir} {\bf 1998}, {\it 14}, 2940.
Sear, R. P. {\it J. Phys. Cond. Mat.} {\bf 1998}, {\it 10}, 1677.

\bibitem{Blankschtein}
Nikas, Y. J.; Blankschtein, D. {\it Langmuir} {\bf 1994}, {\it
10}, 3512.

%\bibitem{grafted}
%Currie, E. P. K.; van der Gucht, J.; Borisov, O. V.; Cohen Stuart,
%M. A. {\it Langmuir} {\bf 1998}, {\it 14}, 5740.

\bibitem{deGennes}
de Gennes, P.-G. {\it J. Phys. (Paris)} {\bf 1976}, {\it 37},
L-59. Brochard, F.; de Gennes, P. G. {\it Ferroelectrics} {\bf
1980}, {\it 30}, 33.

\bibitem{polymer_monolayer}
de Gennes, P.-G. {\it J. Phys. Chem.} {\bf 1990}, {\it 94}, 8407.
Andelman, D.; Joanny, J.-F. {\it J. Phys. II France} {\bf 1993},
{\it 3}, 121. Ch\^{a}tellier, X.; Andelman, D. {\it J. Phys.
Chem.} {\bf 1996}, {\it 100}, 9444.

\bibitem{review_phase}
Piculell, L.; Lindman, B.; Karlstr\"om, G. In
ref~\cite{review_general2}, Chapter 3.

\bibitem{annealed}
This is always the case for annealed impurities, \ie additional
degrees of freedom which are allowed to equilibrate.

\bibitem{review_pe}
Barrat, J.-L.; Joanny, J.-F. {\it Adv. Chem. Phys.}, {\bf 1996},
{\it 94}, 1.

\bibitem{only_ES}
Note that this analysis applies only in the case where $v$ has a
purely electrostatic origin. Results are different when additional
interactions, \eg hydrophobic ones, become significant, as is
demonstrated in Section~\ref{sec_polysoap}.

\bibitem{pe_reservation}
In view of the complexity of polyelectrolyte solutions
\cite{review_pe}, the following, more quantitative attempts
should be considered with caution. Being aware of the limitations
of such calculations, we restrict to the simplest case of
very weak polyelectrolytes.

\bibitem{Fixman}
Fixman, M.; Skolnick, J. {\it Macromolecules} {\bf 1978}, {\it
11}, 863.

\bibitem{c_half}
Many workers have chosen to present experimental results by means
of $c_{1/2}$ (sometimes referred to as $1/Ku$), the surfactant
concentration at which the binding reaches half of its saturation
value, rather than the {\it cac}. The effect of parameters such as
$c_{\rm salt}$ and $I$ on $c_{1/2}$ is a combination of two
different effects: one on the {\it cac} value and another on the
binding cooperativity (isotherm slope). As a result, this
presentation tends to obscure the universal nature of the
phenomenon. In Table~\ref{tab_I} and Figure~\ref{fig_salt} we
present only {\it cac} values, \ie the concentrations
corresponding to the onset of association. Although less sharply
defined, these are the values relevant to our analysis.

\bibitem{Benrraou92}
Benrraou, M.; Zana, R.; Varoqui, R.; Pefferkorn, E. {\it J. Phys.
Chem.} {\bf 1992}, {\it 96}, 1468.

\bibitem{Anthony96}
Anthony, O.; Zana, R. {\it Langmuir} {\bf 1996}, {\it 12}, 1967.

\bibitem{Wei93}
Wei, Y. C.; Hudson, S. M. {\it Macromolecules} {\bf 1993}, {\it
26}, 4151.

\bibitem{Shimizu86}
Shimizu, T.; Seki, M.; Kwak, J. C. T. {\it Colloid Surf.} {\bf
1986}, {\it 20}, 289.

\bibitem{Kiefer93}
Kiefer, J. J.; Somasundaran, P.; Ananthapadmanabhan, K. P. {\it
Langmuir} {\bf 1993}, {\it 9}, 1187.

\bibitem{Chandar88}
Chandar, P.; Somasundaran, P.; Turro, N. J. {\it Macromolecules}
{\bf 1988}, {\it 21}, 950.

\bibitem{Satake90}
Satake, I; Takahashi, T.; Hayakawa, K.; Maeda, T.; Aoyagi, M. {\it
Bull. Chem. Soc. Jpn.} {\bf 1990}, {\it 63}, 926.

\bibitem{Hansson95}
Hansson, P.; Almgren M. {\it J. Phys. Chem.} {\bf 1995}, {\it 99},
16684.

\bibitem{Malovikova84}
Malovikova, A; Hayakawa, K.; Kwak, J. C. T. {\it J. Phys. Chem.}
{\bf 1984}, {\it 88}, 1930.

\bibitem{Hayakawa82}
Hayakawa, K.; Kwak, J. C. T. {\it J. Phys. Chem.} {\bf 1982}, {\it
86}, 3866.

\bibitem{multivalent}
This value seems to be much lower in the case of multivalent salt.
See Hayakawa, K.; Kwak, J. C. T. {\it J. Phys. Chem.} {\bf 1983},
{\it 87}, 506.

\bibitem{review_Ananth}
Ananthapadmanabhan, K. P. In ref~\cite{review_general1}, Chapter
8.

\bibitem{rigid_pe}
Hayakawa, K.; Kwak, J. C. T. In {\it Cationic Surfactants};
Rubingh, D. N.; Holland, P. M., Eds.; Marcel Dekker: New York,
1991.

\bibitem{Wang98}
Wang, G.; Olofsson, G. {\it J. Phys. Chem. B} {\bf 1998}, {\it
102}, 9276.

\bibitem{Corti81}
Corti, M.; Degiorgio, V. {\it J. Phys. Chem.} {\bf 1981}, {\it
85}, 711.

\bibitem{Kamenka97}
Kamenka, N.; Zana, R. {\it J. Colloid Interface Sci.} {\bf 1997},
{\it 188}, 130.

\bibitem{Murata73}
Murata, M.; Arai, H. {\it J. Colloid Interface Sci.} {\bf 1973},
{\it 44}, 475.

%\bibitem{N_strong1}
%J. Liu, N. Takisawa, K. Shirahama, H. Abe and K. Sakamoto,
%{\it J. Phys. Chem. B} {\bf 101}, 7520 (1997).

%\bibitem{N_strong2}
%E. Rodenas and M. L. Sierra,
%{\it Langmuir} {\bf 12}, 1600 (1996).

\bibitem{depend_cmc}
Moreover, according to eq~\ref{cac_general} there should be a weak
decreasing dependence of $\cac/\cmc$ on $\cmc$, which seems to
agree with the data of Table~\ref{tab_weak}.

\bibitem{Sergeyev}
Interestingly, the {\it cac} was found to be accompanied by a
coil-to-globule collapse in certain DNA--surfactant systems:
Mel'nikov, S. M.; Sergeyev, V. G.; Yoshikawa, K. {\it J. Am. Chem.
Soc.} {\bf 1995}, {\it 117}, 2401. {\it ibid.}, 9951.

\bibitem{single_chain}
Note that the following analysis treats a single chain and
is applicable, therefore, only in the dilute polymer limit.

\bibitem{dG_book}
de Gennes, P.-G. {\it Scaling Concepts in Polymer Physics};
Cornell University Press: Ithaca, 1979.

\bibitem{Fisher}
Fisher, M. E. {\it Phys. Rev.} {\bf 1968}, {\it 176}, 257.

\bibitem{Harris}
Harris, A. B., {\it J. Phys. C} {\bf 1974}, {\it 7}, 1671.

\bibitem{d_4}
Note that for a self-avoiding walk with a Flory exponent,
$\nu=3/(d+2)$, the self-consistency condition is satisfied for
$d<4$. This is in accord with the fact that short-range
interactions (in our case the interaction induced by the small
molecules) become irrelevant to polymer statistics for $d\geq 4$.

%\bibitem{finite_xi}
%Unlike the critical system treated by de Gennes and
%Brochard,\cite{deGennes} the system discussed here is not close to a
%critical point and the correlation length ($\xi$) remains finite.
%It is rather the correlation amplitude ($e^2$) which becomes large
%when approaching the {\it cmc}.

\bibitem{no_Fisher}
As the small molecules interacting with
the polymer are allowed to equilibrate (\ie they are represented
by {\it annealed} degrees of freedom),
the Fisher renormalization \cite{Fisher}
seems {\it a priori} to be the more relevant analogy.
Note, however, that
the scaling analysis given above has not included any explicit
annealing. Minimization with respect to $\varphi$ has already
been performed in the calculation
of $v_{\rm ps}$ in Section~\ref{sec_cac} (eq~\ref{veff}).
Once this dependence has
been specified, the partial collapse analysis regards the
correlations (\ie $v_{\rm ps}$) as an effective quenched
perturbation
--- hence the similarity to Harris' analysis. In fact, the
calculation of Appendix~B is analogous to a general
derivation of the Harris criterion; see
Ma, S.-K. {\it Modern Theory of Critical Phenomena};
Benjamin-Cummings: Massachusetts, 1976, Chapter X.
Our self-consistency
condition thereby parallels the requirement that the disorder be a
relevant variable in the renormalization-group sense. 
Indeed, Fisher's change 
of exponents, $\nu\rightarrow\nu/(d\nu-1)$, does not give
relevant results in the current case (\eg for a two-dimensional
polymer with excluded volume, this transformation gives a change
of $\nu$ from 3/4 to the unphysical value of 3/2).
% (\eg $\nu=3/4$ for $d=3$ and $\nu=3/2$ for
%$d=2$).

\bibitem{Nilsson}
Holmberg, C.; Nilsson, S.; Singh, S. K.; Sundel\"of, L.-O. {\it J.
Phys. Chem.} {\bf 1992}, {\it 96}, 871. Nilsson, S. {\it
Macromolecules} {\bf 1995}, {\it 28}, 7837.

\bibitem{Cabane85}
Cabane, B.; Duplessix, R. {\it Colloid Surf.} {\bf 1985}, {\it 13},
19.

\bibitem{Dawson}
Jennings, D. E.; Kuznetsov, Y. A.; Timoshenko, E. G.; Dawson, K. A.
{\it J. Chem. Phys.} {\bf 1998}, {\it 108}, 1702.

\bibitem{Reed}
Norwood, D. P.; Minatti, E.; Reed, W. F. {\it Macromolecules} {\bf
1998}, {\it 31}, 2957. Minatti, E.; Norwood, D. P.; Reed, W. F.
{\it Macromolecules} {\bf 1998}, {\it 31}, 2966.

\bibitem{Gilanyi81}
Gilanyi, T.; Wolfram, E. {\it Colloid Surf.} {\bf 1981}, {\it 3},
181.

\bibitem{persistence}
Ullner, M.; J\"onsson, B; Peterson, C.; Sommelius, O.;
S\"oderberg, B. {\it J. Chem. Phys.} {\bf 1997}, {\it 107}, 1279.
Ha, B. Y.; Thirumalai, D. {\it Macromolecules} {\bf 1995}, {\it
28}, 577. Reed, W. F.; Ghosh, S.; Medjahdi, G.; Francois, J. {\it
Macromolecules} {\bf 1991}, {\it 24}, 6189.

\bibitem{Zana_polysoap}
Binana-Limbel\'{e}, W.;  Zana, R. {\it Macromolecules} {\bf 1987},
{\it 20}, 1331. Binana-Limbel\'{e}, W.;  Zana, R. {\it
Macromolecules} {\bf 1990}, {\it 23}, 2731. Cochin, D.; Candau,
F.; Zana, R. {\it Macromolecules} {\bf 1993}, {\it 26}, 5755. {\it
ibid.}, 5765. Cochin, D.; Candau, F.; Zana, R.; Talmon, Y. {\it
Macromolecules} {\bf 1992}, {\it 25}, 4220. Zana, R.; Kaplun, A.;
Talmon, Y. {\it Langmuir} {\bf 1993}, {\it 9}, 1948. Kamenka, N.;
Kaplun, A.; Talmon, Y.; Zana, R. {\it Langmuir} {\bf 1994}, {\it
10}, 2960. Anthony, O.; Zana, R. {\it Macromolecules} {\bf 1994},
{\it 27}, 3885. Anthony, O.; Zana, R. {\it Langmuir} {\bf 1996},
{\it 12}, 3590.

\bibitem{Chen98}
Chen, L.; Yu, S.; Kagami, Y.; Gong, J.; Osada, Y. {\it
Macromolecules} {\bf 1998}, {\it 31}, 787.

\bibitem{Magny94}
Magny, B.; Iliopoulos, I.; Zana, R.; Audebert, R. {\it Langmuir}
{\bf 1994}, {\it 10}, 3180.

\bibitem{helix}
Satake, I; Gondo, T.; Kimizuka, H. {\it Bull. Chem. Soc. Jpn.}
{\bf 1979}, {\it 52}, 361.

\bibitem{stiff}
Diamant, H.; Andelman, D. {\it Phys. Rev. E} {\bf 2000},
{\it 61}, 6740.

\bibitem{Gaussian} See, \eg Doi, M.; Edwards, S. F. {\it The
Theory of Polymer Dynamics}; Oxford University Press: Oxford,
1986, Chapter 2.

%\bibitem{polysoap}
%M. S. Turner  and J.-F. Joanny,
%{\it J. Phys. Chem.} {\bf 97}, 4825 (1993).
%E. Hamad and S. Qutubuddin,
%{\it Macromolecules} {\bf 23}, 4185 (1990);
%{\it J. Chem. Phys.} {\bf 96}, 6222 (1992).
%O. V. Borisov and A. Halperin,
%{\it Langmuir} {\bf 11}, 2911 (1995);
%{\it Macromolecules} {\bf 29}, 2612 (1996);
%{\it Europhys. Lett.} {\bf 34}, 657 (1996).

\end{thebibliography}
\end{document}